\documentclass[printer]{aa}  

\usepackage{graphicx}
\usepackage{txfonts}
\usepackage{amsmath}
\usepackage{upgreek}
\usepackage{amsfonts}
\usepackage{amssymb}
\usepackage{bbold}
\usepackage{lscape}
\usepackage{tabularx}
\usepackage{supertabular}
\usepackage{xcolor}
\usepackage[normalem]{ulem}
\usepackage{comment}
\usepackage{listings}
\usepackage{lineno}

\newcommand\clearrow{\global\let\rowmac\relax}
\clearrow

\usepackage{hyperref}
\hypersetup{
    colorlinks=true,
    linkcolor=blue,
    citecolor=blue,
    urlcolor =blue
}

\begin{document} 

\title{The preferential orientation of magnetic switchbacks, implications for solar magnetic flux transport}

\author{Naïs Fargette \inst{1}
      \and Benoit Lavraud \inst{1,}\inst{2}
      \and Alexis P. Rouillard \inst{1} 
      \and Victor Réville \inst{1} 
      \and Stuart D. Bale \inst{3}
      \and Justin Kasper \inst{4}
      \fnmsep
      }
\institute{Institut de Recherche en Astrophysique et Planétologie, CNRS, UPS, CNES, Toulouse, France \\
          \and  Laboratoire d'astrophysique de Bordeaux, Univ. Bordeaux, CNRS, Pessac, France
          \and Space Sciences Laboratory, University of California, Berkeley, CA, USA; Physics Department, University of California, Berkeley, CA, USA
          \and Climate and Space Sciences and Engineering, University of Michigan, Ann Arbor, MI, USA
        }

\date{Received March 2022; Accepted May 2022}

\abstract
{Magnetic switchbacks in the solar wind are large deflections of the magnetic field  vector, often reversing its radial component, and associated with a velocity spike consistent with their Alfvénic nature. The Parker Solar Probe (PSP) mission revealed that they were a dominant feature of the near-Sun solar wind. Where and how they are formed remains unclear and subject to discussion.}
{We investigate the orientation of the magnetic field deflections in switchbacks to determine if they are characterised by a possible preferential orientation}
{We compute the deflection angles $\Vec{\uppsi} = [\phi,\theta]^{\mathrm{T}}$ of the magnetic field relative to the theoretical Parker spiral direction for encounters 1 to 9 of the PSP mission. We first characterize the distribution of these deflection angles for quiet solar wind intervals, and assess the precision of the Parker model as a function of distance from the Sun. We then assume that the solar wind is composed of two populations, the background quiet solar wind and the population of switchbacks, characterized by larger fluctuations. We model the total distribution of deflection angles we observe in the solar wind as a weighed sum of two distinct normal distributions, each corresponding to one of the population. We fit the observed data with our model using a Monte-Carlo Markov Chain algorithm and retrieve the most probable mean vector and covariance matrix coefficients of the two Gaussian functions, as well as the population proportion. This method allows us to quantify the properties of both the quiet solar wind and the switchback populations without setting an arbitrary threshold on the magnetic field deflection angles.}
{We first confirm that the Parker spiral is a valid model for quiet solar wind intervals at PSP distances. We observe that the accuracy of the spiral direction in the ecliptic is a function of radial distance, in a manner that is consistent with PSP being near the solar wind acceleration region.
We then find that the fitted switchback population presents a systematic bias in its deflections, with a mean vector consistently shifted toward lower values of $\phi$ ($-5.52^{^\circ}$ on average) and $\theta$ ($-2.15^{^\circ}$ on average) compared to the quiet solar wind population. This results holds for all encounters but E6, and regardless of the magnetic field main polarity. This implies a marked preferential orientation of switchbacks in the clockwise direction in the ecliptic plane, and we discuss this result and its implications in the context of the existing switchback formation theories.
Finally, we report the observation of a 12-hour patch of switchbacks that systematically deflect in the same direction, so that the magnetic field vector tip within the patch deflects and returns to the Parker spiral within a given plane.
}
{}
\keywords{Solar Wind --
            Magnetic Switchbacks --
            Interchange Reconnection --
            Data analysis --
           }
\maketitle

\graphicspath{{./}{figures/}}

\section{Introduction} \label{sec:1_intro}

\textit{Magnetic switchbacks} are structures that are ubiquitous in the near-Sun solar wind and particularly striking in the Parker Solar Probe (PSP) mission data \citep{Kasper_2019, Bale_2019, Horbury_2020, DDW_2020}. Their in-situ signatures include a large deflection of the magnetic field often reversing its radial component - hence their name - associated with a velocity spike consistent with their Alfvénic nature \citep{Matteini_2014, Phan_2020}. In addition, the magnetic field magnitude and the pitch-angle distribution (PAD) of the suprathermal electron population remain fairly constant within switchbacks \citep{Kasper_2019}. From these observations, they are interpreted as large local magnetic folds with faster plasma superposed to a quieter solar wind \citep{Bale_2019}. They had been observed more scarcely in other mission data further away from the Sun \citep{Balogh_1999, Gosling_2011, Horbury_2018} and are now known to be a significant feature of the solar wind below 0.3~AU.

Many physical processes have been proposed to explain the formation of these unexpected structures. One of the most investigated mechanisms is interchange reconnection \citep{Nash_1988, Wang_1989}, where open field lines reconnect with closed ones in the low corona. The foot-point exchange of magnetic field lines provides a theoretical basis to explain how the magnetic field lines can sustain a quasi-rigid rotation in the corona while being anchored in a differentially rotating photosphere \citep{Wang_1996, Fisk_1996, Fisk_1999_APJ}. To keep up with the shear induced by the different rotation rates of the two domains, field lines reconnect at their base and allow coronal hole boundaries to remain unaffected by the photosphere differential rotation \citep{Wang_and_sheeley_2004, Lionello_2005, Lionello_2006}. The newly reconnected magnetic configuration presents a folded magnetic field line, and \cite{Fisk_and_Kasper_2020} proposed that this fold could propagate and become a magnetic switchback at PSP's orbit. However, how such folds could subsist in a low-$\beta$ plasma is unclear, and variation around this mechanism have been proposed. \cite{Zank_2020} argue that interchange reconnection will generate complex structures propagating upward in the solar atmosphere that can reverse their radial field. \cite{Owens_2018, Owens_2020} and \cite{Schwadron_2021} propose that interchange reconnection may lead to a solar wind velocity gradient along open field lines. Subsequently, fast wind overcoming slower wind is able to reverse the magnetic field and create a fold beyond the Alfvén point. \cite{Drake_2021} show through simulation that interchange reconnection can create magnetic flux ropes that present switchback signatures (i.e., radial magnetic field component reversal) and that are very stable and may subsist more easily through propagation in the solar corona and solar wind. \cite{Sterling_2020} investigate coronal jets as a source of switchbacks, arguing that reconnected erupting-minifilament flux rope could generate an Alfvénic fluctuation that steepens during propagation and become a switchback. All of these work assume that switchbacks are created in the low corona through magnetic reconnection. An alternative possibility is that switchbacks could be generated in-situ through processes inherent to solar wind propagation. \cite{Ruffolo_2020} argue that above the Alfvén point, shear-driven dynamics become dominant and accounts for the switchbacks observed by PSP, while \cite{Squire_2020, Shoda_2021, Mallet_2021} link switchbacks to solar wind turbulence. They use compressible MHD simulations and show that expanding Alfvénic fluctuations eventually reverse the magnetic field radial component during propagation. These expanding fluctuations produce magnetic switchback signatures purely born out of turbulence in the solar wind.

The most recent data from PSP brought additional clues to the understanding of switchbacks. An isotropization of the ion distribution function inside switchbacks was observed \citep{Woodham_2021}, showing that plasma properties are different inside switchbacks. They also tend to aggregate in patches \citep{Horbury_2020,DDW_2020} and these patches were found to match the spatial scale of supergranulation \citep{Fargette_2021,Bale_2021}. In addition, switchback patches show an increase in alpha particle abundance compared to the background solar wind \citep{Bale_2021}. All of these recent results indicate that switchback patches, and possibly switchbacks themselves, are distinct from the background solar wind, with different plasma properties, thereby pointing to a formation mechanism in the low corona.

In this work, we investigate over several PSP orbits if the magnetic field deflections display a preferential orientation. \cite{Horbury_2020} performed this type of analysis on a four-day period around the first perihelion of PSP. They report a tendency for long duration switchbacks to deflect in the +T direction of the RTN frame. They also highlight that nearby switchbacks tend to orient themselves in the same direction. A clockwise preference was also observed switchbacks identified in Helios data by \cite{MacNeil_2020}, and the same tendency was identified very recently by \cite{Meng_2022} in encounters 1 and 2 in PSP data.

In section \ref{sec:2_methods}, we present the data analyzed in this work, and detail the methodology and frame we use when defining the switchback phenomenon. In section \ref{sec:3.1}, we characterize the quieter background solar wind and quantify its dispersion around the Parker spiral model. In section \ref{sec:3_orientation}, we model the solar wind as a superposition of a quiet background solar wind and a switchback population. We find that the latter displays a preferential deflection orientation. In section \ref{sec: 4_patch}, we present a particularly striking example of a patch of switchbacks that deflect systematically in the same direction and within the ecliptic plane for 12h. In section \ref{sec:5_discussion}, we discuss our results in the context of the different existing formation theories and discuss implications on solar open flux transport. The conclusions of this study are then given in section \ref{sec:5_conclusion}.

\section{Data and Methods} \label{sec:2_methods}

\subsection{Data}

\begin{figure*}[ht]
    \centering
    \includegraphics[width=1\textwidth]{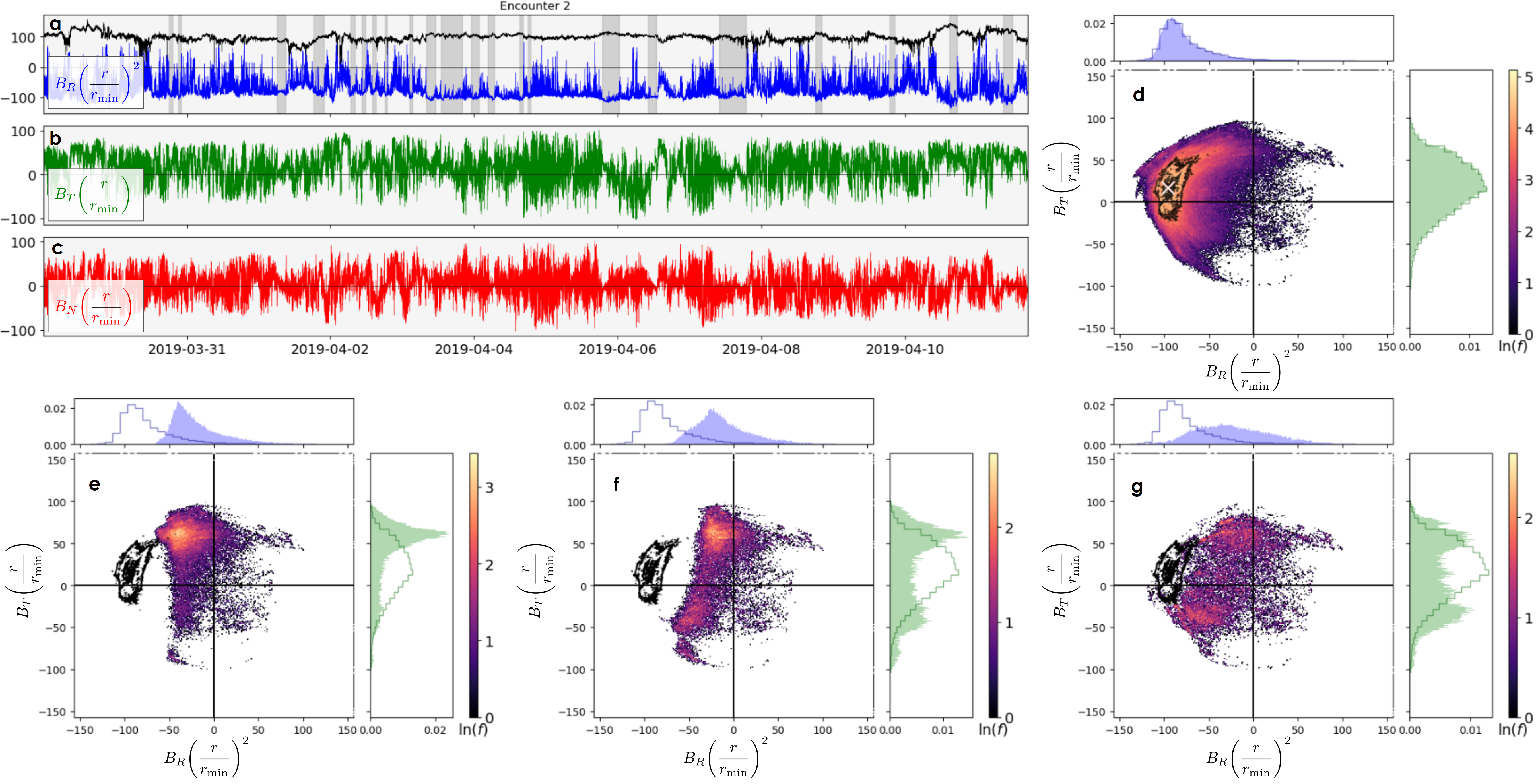}
    \caption{Orientation of the magnetic field for PSP's second encounter. Panels a, b and c we display the magnetic field components normalized by the radial distance $B_R(r/r_{min})^2$, $B_T(r/r_{min})$ and $B_N(r/r_{min})$ (homogeneous to nT), with $r$ the Sun-to-spacecraft distance and $r_{min}= 35.6~R_{\odot}$ the radial distance at perihelion. The normalized magnetic field amplitude $B(r/r_{min})^2$ is displayed as well in black in panel a. The grey shaded intervals are manually selected quiet solar wind intervals (cf section \ref{sec:3.1}). In panel (d) we plot the 2D distribution of $B_R (r/r_{min})^2$ and $B_T (r/r_{min})$ with the marginal distributions on the sides. Black contours surround the core of the distribution and the colorscale represents the number of samples. In the bottom panels, we display the points located more than 60$^o$ away from respectively the radial direction (e), the Parker spiral (f) and a 6h-mode vector (g) with the black contours as a reminder of the core of the total distribution. }
    \label{fig: 1_B_orientation}
\end{figure*}

The PSP mission was launched in August 2018 and is to this date completing its 12$^{th}$ orbit around the Sun. During the past three years the spacecraft gradually scanned the solar wind deeper into the solar corona, as Venus gravity assists brought the perihelion of its highly elliptic orbit closer to the Sun. It reached in turn 35.6~$R_{\odot}$ (0.166 AU, $E_1$ to $E_3$), 27.8~$R_{\odot}$ (0.130 AU, $E_4$ to $E_5$),  20.3~$R_{\odot}$ (0.095 AU, $E_6$ to $E_7$) and 16.0~$R_{\odot}$ (0.074 AU, $E_8$ to $E_9$) where $E_x$ stands for encounter (or orbit)
number $x$.

In this study we analyze magnetic field and particle data taken by the different in-situ instruments. Magnetic field data is provided by the FIELDS instrument suite \citep{Bale_2016} and the particle data by the Solar Wind Electrons Alphas and Protons (SWEAP) instrument suite \citep{2016SSRv..204..131K}. Data from the latter includes  plasma moments from the Solar Probe Cup (SPC) \citep{2020ApJS..246...43C} and plasma moments and electron pitch angle distributions from the Solar Probe ANalyzers (SPANs) \citep{2020ApJS..246...74W, Livi_2021}. All data are re-sampled to a constant time step of 2~seconds and we limit our study to heliocentric radial distances less than 60 $R_{\odot}$. Data is shown in the $RTN$ frame of reference, with $\mathbf{R}$ (radial) being the Sun to spacecraft unit vector, $\mathbf{T}$  (tangential) the cross product between the Sun's spin axis and $\mathbf{R}$, and $\mathbf{N}$ (normal) completes the direct orthogonal frame.

\subsection{Switchback definition}
\label{sec:2.2}
In this work we aim to study statistically the orientation of switchbacks, as described in the introduction (section \ref{sec:1_intro}). However, this poses a difficulty from the start. Indeed, switchbacks are usually identified as a deflection from a background magnetic field, and it is obvious that the choice of this background field will directly affect the results one may obtain on their orientation. In the literature, various background definitions have been used to identify switchbacks in statistical studies, for instance:

\begin{itemize}
    \item The radial direction \citep{Woolley_2020, Wu_2021, Bourouaine_2020,Mozer_2020}
    \item A 6h median field \citep{DDW_2020}
    \item A 6h mean field \citep{Bandyopadhyay_2021}
    \item A 1h mode field \citep{Bale_2019}
    \item A modeled Parker spiral field \citep{Horbury_2020, Laker_2021, Fargette_2021}
\end{itemize}
Various threshold were used from 30 to 90$^o$, as well as additional selection criteria such as duration, field magnitude, Alfvénicity, density, etc., that are not listed here.  Visual selections of switchbacks were also performed often based on radial magnetic field reversals and their duration \citep{Larosa_2021, Martinovic_2021,Akhavan-Tafti_2021}. 

Two kinds of approaches are typically used. One seeks to determine the background magnetic field through post treatment of the data in an attempt to differentiate switchbacks from background solar wind, using different statistical parameters of the magnetic field distribution like mean, median or mode values. The other consists in modeling independently the expected background field using either a radial field assumption or the Parker spiral model. Both methods have their caveats. If the solar wind dynamics is dominated by switchbacks over long periods, as is often the case, then it will be reflected in the mean, median and mode value of the distributions considered, with associated biases. The appropriateness of the modeling approach, on the other hand, will depend on the reliability of the model used and its potential limitations.

The backgrounds obtained through the use of these various methods can be drastically different and lead to different, and sometimes contradictory, results. In a study that will focus on the existence of a preferential orientation within switchback, it is essential to have in mind that this first assumption regarding background modeling may impact the results significantly. In Figure \ref{fig: 1_B_orientation}, we illustrate this fact by comparing the different conclusions one might draw based on such selection processes. We display the magnetic field components measured by PSP in panels a, b and c during encounter 2, from 2019-03-29 00h to 2019-04-11 17h, while the spacecraft was below 60~$R_{\odot}$. The $B_R$ component is normalized by $(r/r_{min})^2$ while $B_T$ and $B_N$ are normalized by $(r/r_{min})$ with $r$ the Sun-to-spacecraft distance and $r_{min}= 35.6~R_{\odot}$ the radial distance at perihelion. PSP is here connected to the negative polarity solar hemisphere throughout the 13 days of data.
In the top right panel (d) we plot the 2D distribution of $B_R (r/r_{min})^2$ and $B_T (r/r_{min})$ with linearly spaced iso-contours (black curves) underlining the core of the distribution. The colorscale represents the number of samples and we also add the normalized projected distributions on the side. In the bottom panels (e to g), we display the points that are located more than 60$^o$ away from respectively the radial direction (e), the Parker spiral (f) and a 6h-mode vector (g) with the black contours as a reminder of the core of the total distribution.

The core of the total distribution (black contours) has a non zero $B_T$ component. This positive $B_T$ component is consistent with the Parker spiral. Indeed, when we take the average of the Parker spiral angle throughout the encounter (by taking the mean of equation \eqref{eq: spiral} over E2, see section \ref{sec:2.3}), we obtain an angle of $167 \pm3~^o$ away from the radial direction. This corresponds to $B_T (r/r_{min}) = 17\pm3$~nT and $B_R (r/r_{min})^2 = -96\pm13$~nT, and we mark these values with a white cross in panel d for illustrative purposes. In the following panels (e to g) it is clear that the distributions obtained through the three methods differ significantly. With the radial method, the $B_T$ component of a modeled Parker spiral is neglected and as a direct consequence the deviation one detects will be strongly biased toward a positive $B_T$. By contrast the distribution 60$^{\circ}$ away from the Parker spiral includes more points with a negative $B_T$ while keeping a preference toward a positive $B_T$. Finally, when we set the switchback definition 60$^o$ away from a sliding mode, the tangential distribution of the magnetic field is even more isotropic. From these plots, it is clear that if we want to investigate a possible systematic orientation, we cannot define switchbacks based solely on the radial direction because the tangential component of the Parker spiral is significant.

Defining switchbacks as a perturbation relative to the Parker spiral appears as the most physically motivated approach for our purpose. In Figure \ref{fig: 1_B_orientation}d, we see that the spiral accurately models the core of the magnetic field orientation distribution. To study deviations compared to a median or mode field may also be useful is some contexts but calls for a different interpretation of the results, as one then studies rapid fluctuations compared to slower fluctuations of the field. 
In this work we choose the Parker spiral as the modeled background field and check for its accuracy before analysing the switchback perturbation.

\subsection{Coordinate system}
\label{sec:2.3}
The Parker spiral angle is the trigonometric angle between the radial direction and the spiral direction in the RT plane, given by \citep{Parker_58}:

\begin{equation}
    \alpha_p(t) = \arctan2 \left( \dfrac{-\omega \left(r(t)-r_0\right)} {V_r(t)}\right)
    \label{eq: spiral}
\end{equation}

where $\omega = 2.9\times10^{-6}$~s$^{-1}$ is the Sun's rotational frequency taken at the equator, $r(t)$ is the distance of the spacecraft to the center of the Sun, $r_0= 10~R_{\odot}$ \citep{Bruno_97} is the source distance of the Parker spiral, and $V_r(t)$ is the measured radial speed of the solar wind. For our purpose, we use the velocity processed with a low pass filter characterized by a cutting wavelength at 2h. This allows for the removal of spurious data, as well as short timescale variations and transient structures that are not relevant to the Parker spiral angle. We thus obtain a Parker spiral angle varying over time with a timestep of 2~s, similarly to other quantities.

To compare the magnetic field orientation to the expected local Parker spiral calculated with $r$ and $V_r$, we transform each data point into its local Parker frame $\mathbf{x}, \mathbf{y}, \mathbf{z}$ where $\mathbf{x}$ points in the direction of the local spiral, $\mathbf{z}=\mathbf{N}$ remains unchanged from the RTN frame, and $\mathbf{y}$ completes the direct orthogonal frame. An important point is that this frame rotates as a function of the polarity of the solar magnetic field, and a magnetic field matching the local spiral perfectly is then written as $\mathbf{B} = B_x \mathbf{x}$ with $B_x$ positive. Finally when studying orientation, it is convenient to use a spherical coordinate system $(||\mathbf{B}||, \phi, \theta)$, where $\phi$ and $\theta$ are the azimuthal and elevation angle in this $xyz$ Parker frame spanning respectively $[-180,180]^{\circ}$ and $[-90, 90]^{\circ}$.  
We will hereafter write $\Vec{\uppsi} = [\phi, \theta]^\mathsf{T}$ the vector containing the orientation angles of the magnetic field.

\section{quiet solar wind orientation} 
\label{sec:3.1}

\begin{figure}[h]
    \centering
    \includegraphics[width=.49\textwidth]{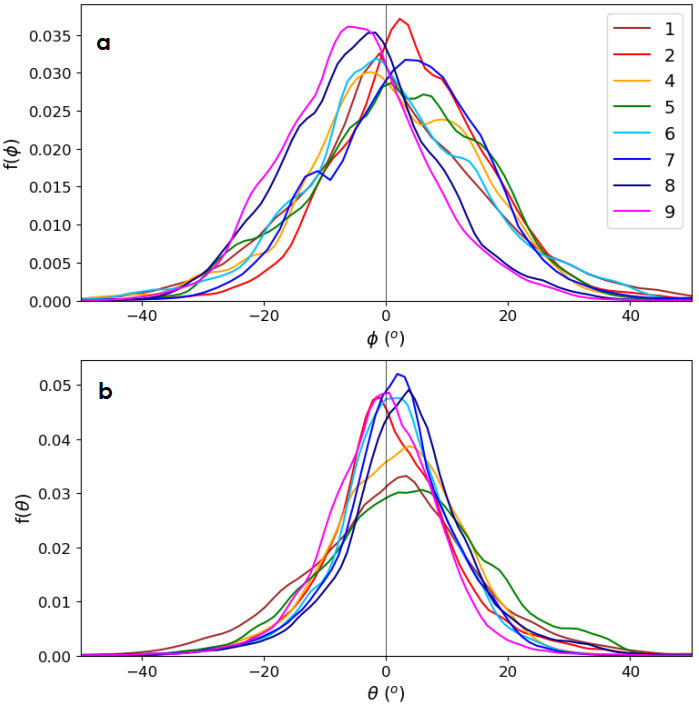}
    \caption{Distributions of orientation angles $\phi$ (panel a) and $\theta$ (panel b) for quiet solar wind intervals over encounters 1 to 9.}
    \label{fig: 2_calm_wind}
\end{figure}

The first step of our study is to quantify the accuracy of the Parker model that we want to use for the background field. To do so, we manually selected periods of quiet solar wind in the time series as periods that were not dominated by large scale fluctuations. We chose periods that lasted at least one hour with no or very few deviations greater than $60^{\circ}$ from the expected spiral direction. This selection was performed visually which can lead to a selection bias despite our best efforts. We hence give the timetable of the selected intervals in Appendix \ref{sec: app_A}, and the grey shaded intervals in Figure \ref{fig: 1_B_orientation} illustrate these for encounter 2. In Figure \ref{fig: 2_calm_wind}, we display the distribution of the orientation angles $\phi$ and $\theta$ inside these quiet solar wind intervals, with the colors differentiating the different encounters.

\begin{table}[h]
    \centering
    \caption{Median vectors and associated dispersion of the quiet solar wind distributions displayed in Figure \ref{fig: 2_calm_wind}}
    \label{tab:calm_wind}
    \begin{tabular}{l|c|c}
        Enc & \Vec{\uppsi} Median ($^{\circ}$) & \Vec{\uppsi} Dispersion ($^{\circ}$) \\
        \hline \hline
		1 & [0.4 , 1.2] & [16.6 , 14.6] \\
		2 & [4.0 , 0.5] & [12.7 , 11.2] \\
		4 & [0.8 , 2.1] & [14.3 , 10.8] \\
		5 & [2.4 , 3.7] & [14.5 , 13.9] \\
		6 & [1.1 , 1.7] & [14.8 , 9.7] \\
		7 & [3.7 , 1.8] & [13.4 , 10.2] \\
		8 & [-3.9 , 3.3] & [12.5 , 10.7] \\
		9 & [-5.3 , -0.4] & [12.0 , 9.7] \\
		\hline
    \end{tabular}
\end{table}

\begin{figure}[h]
    \centering
    \includegraphics[width=.49\textwidth]{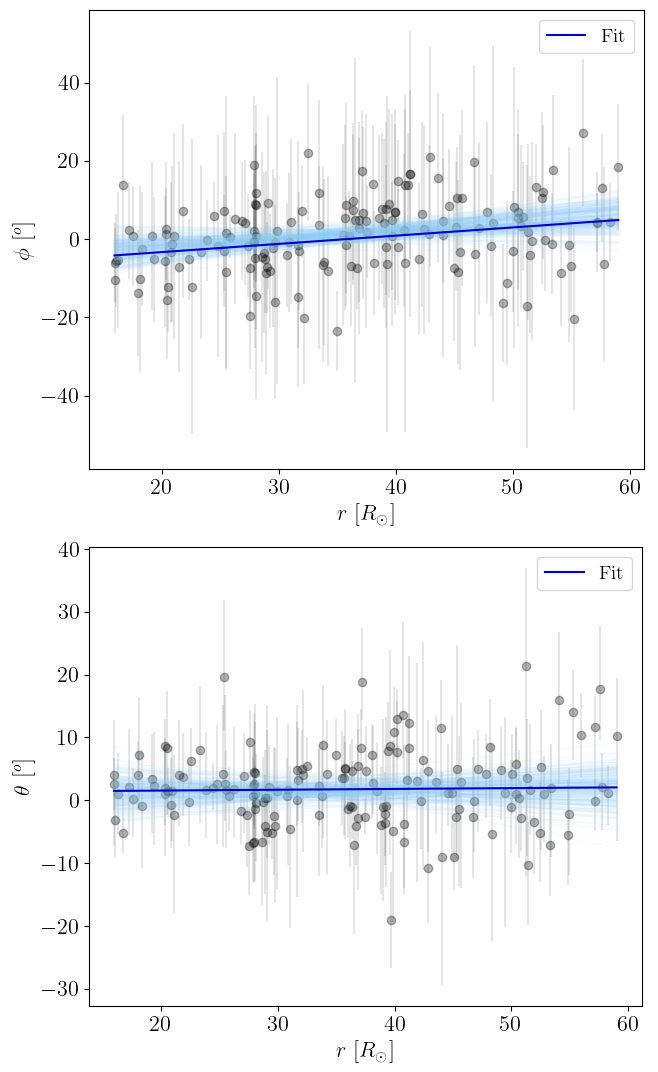}
    \caption{Orientation angles of the quiet solar wind intervals given in table \ref{tab:calm_wind} as a function of radial distance to the Sun. The median value of each interval is plotted (grey dots) with the dispersion inside the interval (error bar).}
    \label{fig: 21_angles_r}
\end{figure}

\begin{figure*}[ht]
    \centering
    \includegraphics[width=1\textwidth]{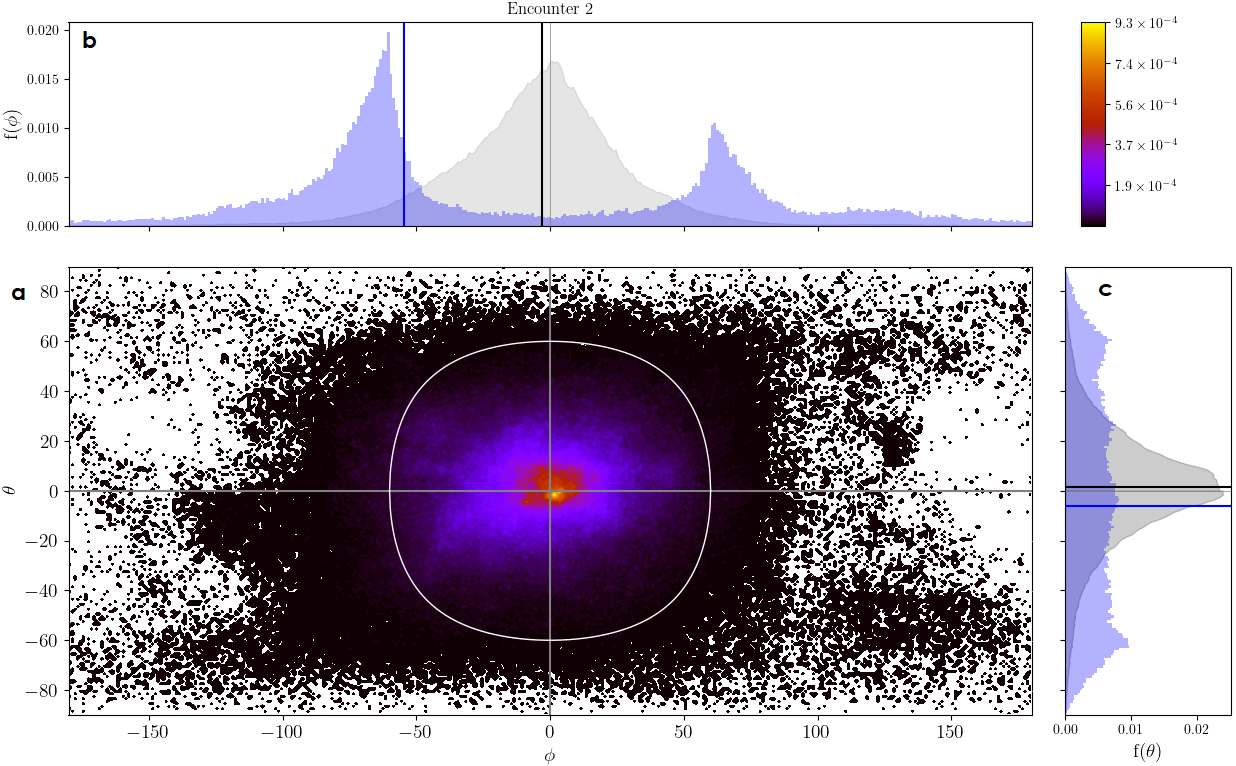}
    \caption{2D normalized distribution of magnetic field orientation angles for encounter 2 (panel a) together with the marginal distributions of $\phi$ and $\theta$ in light grey in panels b and c. Black lines indicate the median values of the marginal distributions. The white line in panel a corresponds to a $60^{\circ}$ threshold angle (cf text for more details) and in panels b and c we overlay in light blue the distribution of the points outside this line. The blue lines indicate the median values of these truncated blue  distributions.}
    \label{fig: 3_E2_distribution}
\end{figure*}
The magnetic field orientation in these quiet solar wind intervals matches reasonably well the Parker spiral direction given by $\Vec{\uppsi} = [0^{\circ},0^{\circ}]^\mathsf{T}$. \footnote{To give the reader a range of comparison, this $\Vec{\uppsi} = [0^{\circ},0^{\circ}]^\mathsf{T}$ direction corresponds to angles relative to the radial direction between 5.6$^{\circ}$ and 29.6$^{\circ}$  depending on $r$ and $V_r$.} The statistical parameters of the distributions are given in table \ref{tab:calm_wind}, with on average a median vector of $[0.4^{\circ}, 1.8^{\circ}]^{^\mathsf{T}}$ and associated standard deviations of $[13.9^{\circ}, 11.3^{\circ}]^{^\mathsf{T}}$. Interestingly, we note a tendency for encounters 8 and 9 to have a median value and peak biased toward negative $\phi$. We investigated if this could be due to PSP approaching closer to the Sun for the latest encounters. In Figure \ref{fig: 21_angles_r}, we plot for each quiet solar wind interval the median  orientation of the angles $\phi$ (panel \ref{fig: 21_angles_r}a) and $\theta$ (panel \ref{fig: 21_angles_r}b)  as a function of the spacecraft distance $r$ (grey dots), and add the associated standard deviation (grey bars). We find a Spearman correlation coefficient (measuring the degree of monotonicity between two variables) of 0.28 for $(\phi,r)$ with an associated p-value of $3.10^{-4}$, and of 0.05 for $(\theta,r)$ with an associated p-value of $0.5$. Eventhough the correlation coefficient of $\phi$ with $r$ is low, the small p-value indicates that the probability of observing such a dataset with randomly distributed variables is of $3.10^{-4}$ (and as such, unlikely). This shows that although weak, the correlation between $\phi$ and $r$ seems significant, while that between $\theta$ and $r$ does not. We also fitted a linear model to the data and found that $\phi =0.235^{+0.081}_{-0.128} r - 6.0^{+3.3}_{-4.4}$ and $\theta = 0.049^{+0.072}_{-0.122} r + 0.4^{+3.8}_{-3.3}$, once again confirming that $\phi$ slightly increases with distance $r$. The fits are shown in blue in Figure \ref{fig: 21_angles_r} with light blue curves indicating the uncertainty of the fit. This does not necessarily mean that the relation between the two variables is linear, as indeed the increase is mainly visible below $30~R_{\odot}$ and in data from E8 and E9. This result should be confirmed by measurements from further encounters, nonetheless we discuss its implications in section \ref{sec:5.1}.

\section{Global orientation} \label{sec:3_orientation}

 \subsection{Modeling switchbacks}
 \label{sec:3.2}

We now consider the complete 2D distribution of magnetic field orientation angles for encounter 2 $f(\Vec{\uppsi})$, spanning 13 days of data with a 2 second timestep. It is displayed in Figure \ref{fig: 3_E2_distribution}a together with the marginal (i.e. projected) distributions of $\phi$ (\ref{fig: 3_E2_distribution}b) and $\theta$ (\ref{fig: 3_E2_distribution}c). The distribution is characterized by a  median vector of $\Vec{\uppsi} = [-2.9^{\circ}, 1.4^{\circ}]^\mathsf{T}$ (black lines in (\ref{fig: 3_E2_distribution}b and c) with associated standard deviation of $[34.7^{\circ},22.6^{\circ}]^\mathsf{T}$, hence wider than the quiet solar wind distribution and consistent with the presence of a population of larger fluctuations. We can see that the peak of the distribution remains around $[0^{\circ},0^{\circ}]^\mathsf{T}$ as it was for the quiet solar wind.

\begin{figure*}[ht]
    \centering
    \includegraphics[width=1\textwidth]{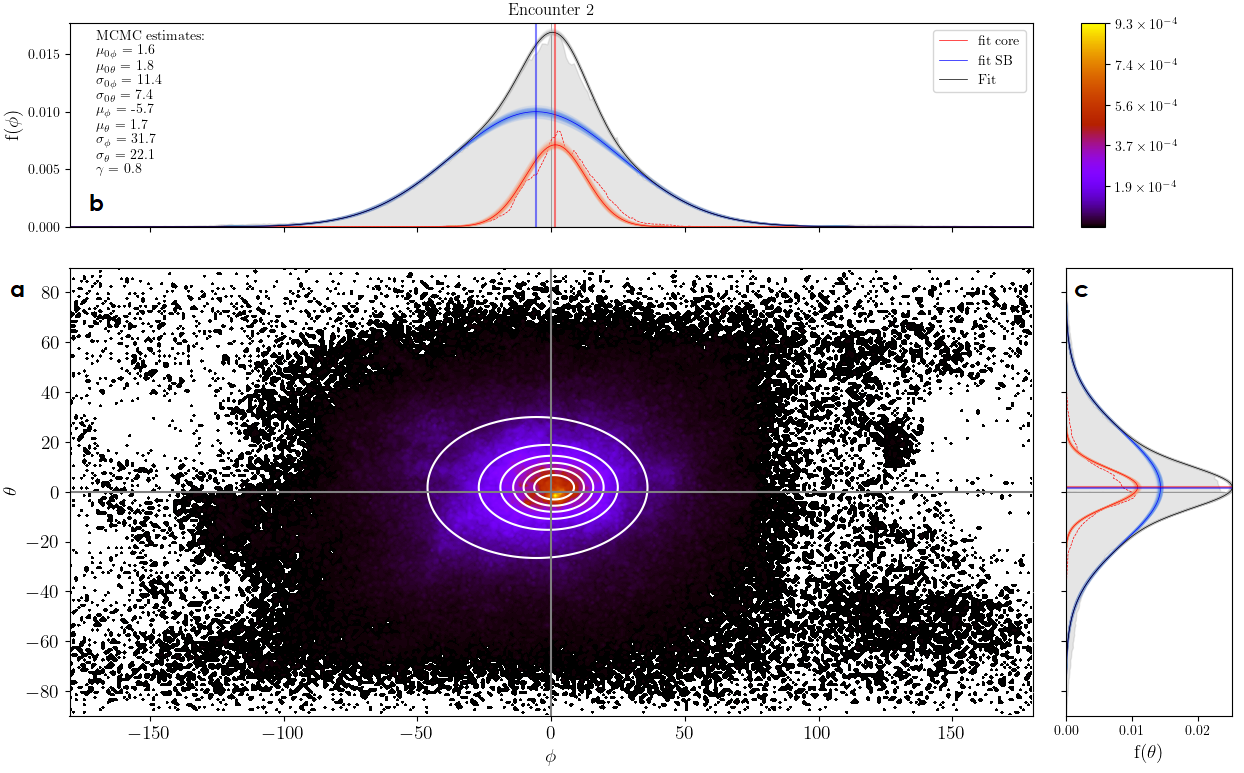}
    \caption{2D distribution of magnetic field orientation angles for encounter 2 (panel a) together with the marginal distributions of $\phi$ and $\theta$ in light grey in panels b and c. The white contours (a) represent the fitted function, and its marginal distributions are in black in panel b and c. We also plot in panels b and c the marginal distributions corresponding to quiet solar wind (red) and switchback (blue) populations, with lines indicating their mean. The curves in lighter red, blue and black give a sense of the fit precision. Finally in dashed red we display the quiet solar wind distribution found for E2 as displayed in Figure \ref{fig: 2_calm_wind} and multiplied by $(1-\gamma)$ so that the scales are comparable. See text for more detail.}
    \label{fig: 4_fits}
\end{figure*}

A usual method chosen to study switchbacks is to segregate the two populations - background wind and switchbacks - based on a chosen threshold angle. Given the quiet solar wind distribution displayed in Figure \ref{fig: 2_calm_wind} we find that this threshold should be taken at a minimum of around $40^{\circ}$ (three standard deviation away) in the $\phi$ direction. This threshold is usually taken on the angle between $\mathbf{B}$ and $\mathbf{x}$ (with $\mathbf{x}$ the unit vector of the Parker spiral, see section \ref{sec:2.3}), that is linked to $\phi$ and $\theta$ through $\mathbf{b} .\mathbf{x} = \cos \theta \cos \phi$ with $\mathbf{b}$ the unit vector of $\mathbf{B}$. In  panel \ref{fig: 3_E2_distribution}a we draw the limit corresponding to a $60^{\circ}$ threshold angle, characterized by $\cos \theta = (2 \cos \phi )^{-1}$, and we overlay in light blue the distribution of the points outside this limit in panels b and c. These points are characterized by a median vector of $[-54.7^{\circ}, -6.2^{\circ}]^{\mathsf{T}}$ (blue lines in (\ref{fig: 3_E2_distribution}b and c) and with associated standard deviations $[79.6^{\circ}, 43.4^{\circ}]^{^\mathsf{T}}$. We notice that large-scale fluctuations occur in all directions around the Parker spiral angle and that their distribution is biased toward negative values of $\phi$ and $\theta$, which correspond to the +T and -N directions in a magnetic field of negative polarity. By construction, in this threshold approach the switchback distribution (in blue) is a truncated distribution. 
In the rest of the analysis we adopted a more continuous probabilistic approach considering the superposition of two solar wind populations with distinct normal distributions in deflection angles. Importantly, we underline here already that both methods (segregation or mixing and fitting of two populations) find consistent results in terms of switchback preferential direction.

For the second approach we assume that the wind is composed of two populations with distinct distribution properties, respectively representing the background, quiet solar wind and the population of switchbacks characterized by larger fluctuations. In accordance with the results of section \ref{sec:3.1} we assume that the quiet solar wind magnetic field deflections follow a 2D normal distribution $\mathcal{N}(\Vec{\upmu_0}, \Sigma_0)$ that should remain close to the Parker spiral, together with a superposed second population of larger deflections $\mathcal{N}(\Vec{\upmu}, \Sigma)$ representing the switchbacks, where $\Vec{\upmu}$ and $\Sigma$ are respectively the mean vector and the covariance matrix of the considered distributions. The total distribution we observe in Figure \ref{fig: 3_E2_distribution} can then be modeled by the sum of the two normal distributions, weighted with a given proportion $\gamma$. This model is written :
\begin{equation}
    f_m(\Vec{\uppsi}, \mathbf{P}) = (1-\gamma)~\mathcal{G}(\Vec{\uppsi},\Vec{\upmu_0}, \Sigma_0) +\gamma~ \mathcal{G}(\Vec{\uppsi},\Vec{\upmu}, \Sigma)
\end{equation}
where  $f_m$ is the modeled distribution, $\mathcal{G}(\Vec{\uppsi},\Vec{\upmu_0}, \Sigma_0)$ and $\mathcal{G}(\Vec{\uppsi},\Vec{\upmu}, \Sigma)$ are 2D Gaussian functions of respective mean vectors $\Vec{\upmu_0} = [\mu_{0\phi}, \mu_{0\theta}]^\mathsf{T}$, $\Vec{\upmu} = [\mu_{\phi}, \mu_{\theta}]^\mathsf{T}$ and covariance matrices $\Sigma_0 = \mathrm{diag}~ (\sigma_{0\phi},\sigma_{0\theta})$,  $\Sigma = \mathrm{diag}~ (\sigma_{\phi},\sigma_{\theta})$, and finally $\mathbf{P}$ is the parameter vector to fit containing 9 parameters :
\begin{equation}
    \mathbf{P} = 
    \begin{bmatrix}
    \mu_{0\phi} & \mu_{0\theta} & \sigma_{0\phi} & \sigma_{0\theta} & \mu_{\phi} &\mu_{\theta} &  \sigma_{\phi} & \sigma_{\theta} & \gamma\\\end{bmatrix}^\mathsf{T}
\end{equation}

We assume that our data, i.e., the distribution $f$, follows our model $f_m$ with a white noise model, and we take the associated dispersion $\sigma_{\epsilon}$ to be 10\% of the maximum of $f$.
From here for a given set of parameter $\Vec{\mathrm{P}}$, the likelihood of the data will follow a 2D normal distribution and may be written: 
\begin{equation}
    \mathrm{p}(f~|~\Vec{\uppsi}, \Vec{\mathrm{P}}) = \mathcal{G}\left(f, ~f_m(\Vec{\uppsi}, \Vec{\mathrm{P}}),~ \sigma_{\epsilon} \mathbb{1} \right)
\end{equation}
where $\mathrm{p}(X)$ designates the probability of X and $\mathbb{1}$ is the identity matrix. We use uniform priors $\mathrm{p}(\Vec{\mathrm{P}})$ on all of the parameters, with the constraints $\mu_{0\phi}, \mu_{0\theta}\in [-10^{\circ},10^{\circ}]$, $ \sigma_{0\phi}, \sigma_{0\theta}\in [0.1^{\circ},30^{\circ}]$. These constraints are based on the results from section 2 where we found a mean close to zero and dispersion of around 15$^{\circ}$.

We can now find the most probable parameters to fit our distribution, and hence seek to maximize the log-posterior probability of the model through the Bayes equation:
\begin{equation}
\ln \mathrm{p}(\Vec{\mathrm{P}}~|~\Vec{\uppsi},f) = \ln \mathrm{p}(\Vec{\mathrm{P}}) + \ln \mathrm{p}(f~|~\Vec{\uppsi}, \Vec{\mathrm{P}}) + \mathrm{C}
\end{equation}
where $\mathrm{C}$ is a constant.

We sample the parameter space using the \textit{emcee} python library \citep{Foreman-Mackey_2019} which is based on a Monte-Carlo Markov chain algorithm, using 32 walkers and 2000 iterations. In Appendix \ref{sec: app_B}, we display the convergence of the 32 walkers over the 2000 iterations (Figure \ref{fig: app_fourmis}), and show the probability distribution function of the walker positions in the 9D space, discarding the first 1000 iterations (Figure \ref{fig: app_mcmc}). This yields the most probable parameter vector $\Vec{\mathrm{P}}$ which is summarized in table \ref{tab:model_params}.
\begin{table}
    \centering
    \caption{Most probable (maximum a-posteriori) parameter vector $\Vec{\mathrm{P}}$, obtained after fitting the double Gaussian model described in the text to the data from encounter 2. The first line presents the parameters associated with the background quiet solar wind model, the second line those associated with the switchback population, and on the third line is the proportion of switchback population.}
    \label{tab:model_params}
    \begin{tabular}{cccc}
        \hline \hline
		$\mu_{0\phi}$ & $\mu_{0\theta}$ & $\sigma_{0\phi}$ & $\sigma_{0\theta}$ \\ 
		\hline
		&&& \\
		$1.61^{+0.23}_{-0.18}$ & $1.81^{+0.14}_{-0.12}$ & $11.32^{+0.32}_{-0.28}$ & $7.44^{+0.15}_{-0.19}$ \\ 
		&&& \\
		\hline \hline
		$\mu_{\phi}$ &$\mu_{\theta}$ &  $\sigma_{\phi}$ & $\sigma_{\theta}$ \\ 
		\hline
		&&& \\
		$-5.68^{+0.42}_{-0.48}$ & $1.71^{+0.29}_{-0.25}$ & $31.75^{+0.43}_{-0.44}$ & $22.13^{+0.29}_{-0.42}$ \\ 
		&&& \\
		\hline \hline
		$\gamma$ & & & \\
		\hline
		&&& \\
		$0.7975^{+0.0095}_{-0.0094} $ & & & \\
    \end{tabular}
\end{table}

In Figure \ref{fig: 4_fits} we present the 2D distribution of the magnetic deflection angles in the same manner as in Figure \ref{fig: 3_E2_distribution}, together with the fitting result. The white contours in panel \ref{fig: 4_fits}a represent the fitted function corresponding to the parameters in table \ref{tab:model_params}, together with the marginal distribution in black in panel b and c. We also plot in panels b and c the projected quiet solar wind distribution $(1-\gamma)*\mathcal{G}(\Vec{\uppsi}, \Vec{\upmu_0}, \Sigma_0)$ in red and the switchback distribution $\gamma* \mathcal{G}(\Vec{\uppsi}, \Vec{\upmu}, \Sigma)$ in blue. To give a sense of the fit precision we also plot in lighter red, blue and black a hundred similar functions with parameters drawn randomly from the parameter probability distribution displayed in the Appendix (Figure \ref{fig: app_mcmc}). Finally in dashed red we display the quiet solar wind distribution found for E2 as displayed in Figure \ref{fig: 2_calm_wind} and multiplied by $(1-\gamma)$ so that the scales are comparable.

What is striking is first that the fitted function (in black in panels b and c) follows quite well the 2D data distribution, and second that the fitting algorithm finds a Parker spiral distribution (in red in panels b and c) with characteristics very similar to the one found in section 2 in an independent manner (see table \ref{tab:calm_wind} line 2 and table \ref{tab:model_params} line 1). We can see that as expected, the switchback population in blue presents a larger dispersion in both dimensions. Its mean vector, however, is different from that of the quiet solar wind population in red. It presents a negative value in the $\phi$ dimension  $\upmu_{\phi} = -5.68^{+0.42}_{-0.48}$. This negative $\upmu_{\phi}$ is consistent with the result found with the previous method (Figure \ref{fig: 3_E2_distribution}) when we considered the median of points with a large deviation from the Parker spiral. In the $\theta$ dimension, however, we find no difference between the means of the core and switchback population while in the previous method we had found a slight tendency toward negative $\theta$. This discrepancy can be explained by the fact that the tail of the marginal distribution in negative $\theta$  is not well reproduced by the fit (panel \ref{fig: 4_fits}c).
Finally we find a proportion of switchback population close to 80\%. This high proportion is due to the method we are using, and can be interpreted as the proportion of the observed solar wind that is dominated by magnetic switchbacks.

\begin{figure}[ht]
    \centering
    \includegraphics[width=.5\textwidth]{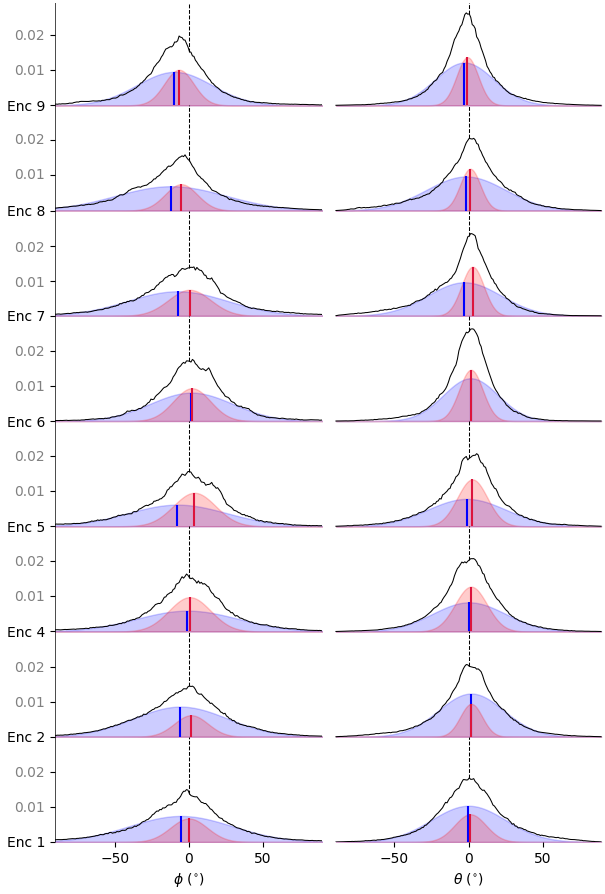}
    \caption{The marginal distributions of observed magnetic field orientation angles $f(\phi)$ (a) and $f(\theta)$ (b) are displayed in black for all encounters, with a shared y-axis. The fitting result are also plotted : in light red the marginal distributions corresponding to quiet solar wind, and in light blue the marginal distribution corresponding to switchbacks. Vertical lines lines indicate their mean, and the dashed line in the background is the zero value.}
    \label{fig: 6_marginals}
\end{figure}

To summarize, in our method we assume that the solar wind magnetic field fluctuations are composed of two populations, each with orientation angles that follow a 2D normal distribution. The first is assumed to follow the Parker spiral with a rather small dispersion, and the second is the switchback population with a wider dispersion. After fitting this model to our data, we find that the background population we retrieve is consistent with the quiet solar wind distribution described in section \ref{sec:3.1}. We also find that the switchback population is biased with an offset in the $-\phi$ direction.These results are confirmed with the more simple analysis of Figure \ref{fig: 3_E2_distribution}, looking at the median values of points more than $60^{\circ}$ away from the spiral, which also shows a preferential $-\phi$ orientation.

\subsection{A systematic bias in the deflections}

We now apply the same method to the remaining encounters. For each, we consider the available data below 60 $R_{\odot}$, and we discard intervals where the Parker spiral model is not relevant, i.e where we identified Heliospheric Current Sheet (HCS) crossings, Coronal Mass Ejections (CMEs) or Flux Ropes (FR). This selection was done manually by analysing the magnetic field, plasma moments and the pitch angle distribution of suprathermal electrons; it can be reviewed in Appendix \ref{sec: app_C}. In order to identify a potential influence of the magnetic field polarity, we also restrained our study to the main polarity of each encounters. It means that we considered only the data points when the spacecraft was sampling a negative polarity solar wind (south of the HCS) for encounters 1, 2, 4, 5, 6 and 9, and a positive polarity solar wind (north of the HCS) for encounter 7 and 8. We compute the orientation angles of the magnetic field in the local Parker frame and fit the obtained distribution for the most probable parameters in the same manner as in Table \ref{tab:model_params}. 

\begin{figure}[ht]
    \centering
    \includegraphics[width=.48\textwidth]{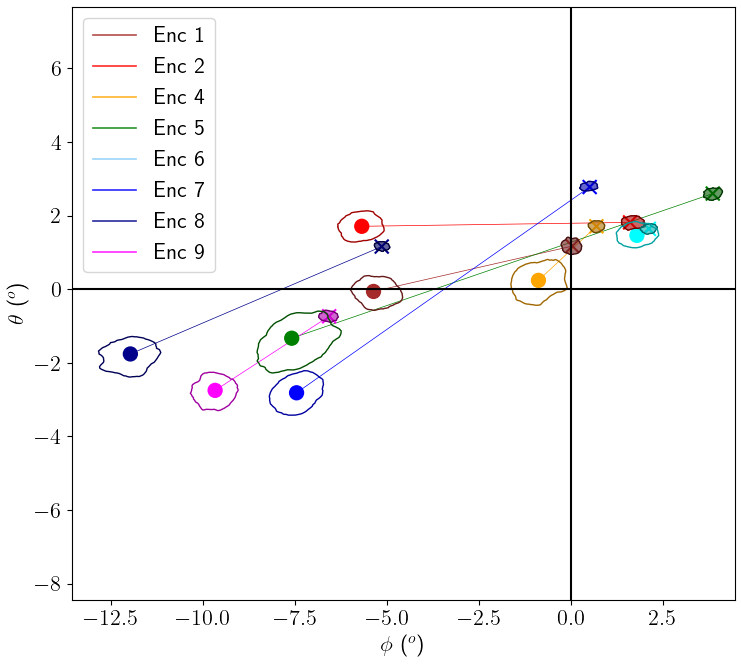}
    \caption{Mean vectors of quiet solar wind (cross) and switchback (dot) population for each encounter. Both are linked by a line for visualization purposes. Contours around the markers (filled for quiet solar wind, transparent for switchbacks) indicate the uncertainty of the fit we performed.}
    \label{fig: 6_bias}
\end{figure}

\begin{table}[h]
    \centering
    \begin{tabular}{c|c|c}
        Enc & $\Delta \mu_{\phi}$ ($^{\circ}$) & $\Delta \mu_{\theta}$ ($^{\circ}$) \\
        \hline \hline
		1 & -5.41 & -1.22 \\
		2 & -7.25 & -0.11 \\
		4 & -1.60 & -1.38 \\
		5 & -11.47 & -3.81 \\
		6 & -0.37 & -0.18 \\
		7 & -8.06 & -5.58 \\
		8 & -6.86 & -2.87 \\
		9 & -3.12 & -2.02 \\
		\hline
		$\langle\cdot\rangle$  & -5.52 &  -2.15\\ 
		\hline
    \end{tabular}
    \caption{Shift between the quiet solar wind and switchback distribution means. The last line is the average over all encounters}
    \label{tab:shift}
\end{table}

The results are summarized in Figures \ref{fig: 6_marginals} and \ref{fig: 6_bias} and fully available in Appendix \ref{sec: app_B}. 
In Figure \ref{fig: 6_marginals}, we display all the fits we performed for the different encounters by looking at the marginal distributions. We plot in black $f(\phi)$ and $\theta$, corresponding to the marginal distributions of the magnetic field orientation observed by PSP for each encounter. The filled curves are the fitted marginal distributions of the quiet solar wind (in red) and the switchbacks (in blue) with vertical colored lines indicating their mean value. We note that the plots shown for encounter 2 in Figure \ref{fig: 6_marginals} are the same as the ones detailed in Figure \ref{fig: 4_fits}b and \ref{fig: 4_fits}c. This visualization shows that, to first order, the data is accurately reproduced by the model we use, i.e. the weighed superposition of two Gaussian functions. We also see that the switchback distribution (in blue) clearly shows a biased mean shifted toward smaller values of $\phi$ (and $\theta$ to a smaller degree) compared to the quiet solar wind, for all encounters independently apart from E6.

\begin{figure*}[ht]
    \centering
    \includegraphics[width=1\textwidth]{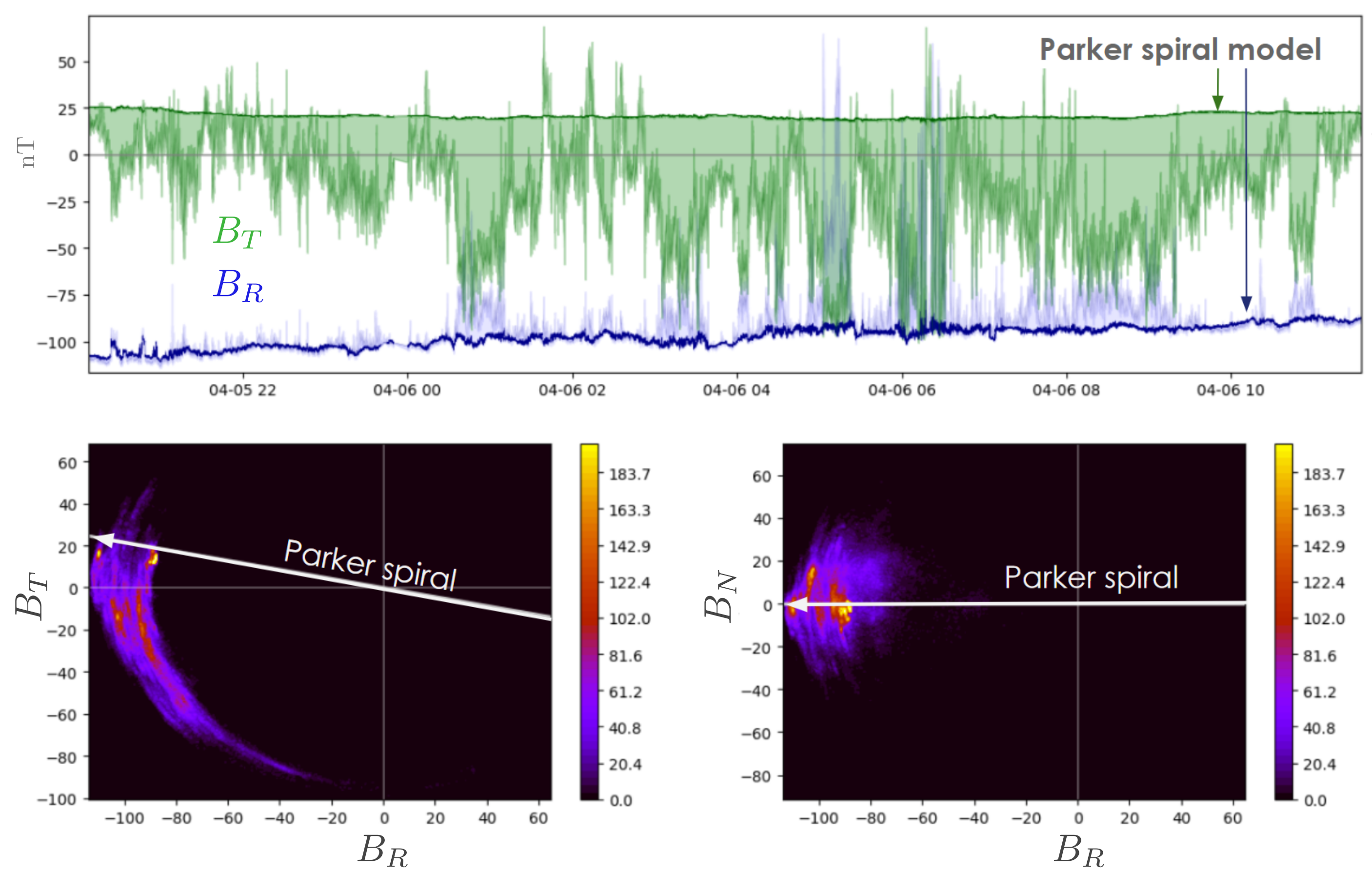}
    \caption{Illustration of a one-sided patch of switchbacks. The top panel shows the radial and tangential component of the magnetic field $B_R$ and $B_T$, as well as the expected components of the Parker spiral magnetic field. The difference between model and data is lightly shaded. In the bottom panels, we plot the 2D distribution $f(B_R, B_T)$ and $f(B_R, B_N)$, the color scale indicates the number of points inside each bin.}
    \label{fig: patch}
\end{figure*}

In Figure \ref{fig: 6_bias}, we display a scatter plot of the mean vector of each population for all encounters, each displayed with a different color. The cross markers indicate the mean vectors found for the quiet solar wind population $\Vec{\upmu_0} = [\mu_{0\phi}, \mu_{0\theta}]^\mathsf{T}$ (corresponding to the red vertical lines in Figure \ref{fig: 6_marginals}), while the filled circles are the mean vectors found for the switchback population $\Vec{\upmu} = [\mu_{\phi}, \mu_{\theta}]^\mathsf{T}$ (vertical blue lines in Figure \ref{fig: 6_marginals}). Each couple of points is linked by a line for visualization purposes. Finally, contours around the markers (filled for quiet solar wind, transparent for switchbacks) indicate the uncertainty (1$\sigma$) of the fit we performed. 
For E8 and E9 the quiet solar wind means have negative $\phi$ values, which is consistent with the values presented in section \ref{fig: 2_calm_wind}. The shifts between the means of the quiet solar wind and switchback distributions are given for each encounter in table \ref{tab:shift}, with $\Delta \mu_{\phi}= -5.52~^{\circ}$ and $\Delta \mu_{\theta} = -2.15^{\circ}$ on average. For all encounters except E6, the switchback population is shifted significantly to lower values of $\phi$, while for all encounters except E2 and E6, it is also somewhat shifted toward lower values of $\theta$, even though the trend is less significant. These results are further discussed in section \ref{sec:5.2}.

\section{Case study of a unidirectional planar patch of switchbacks}
\label{sec: 4_patch}

In addition to the large scale systematic bias found over the different encounters (section \ref{sec:3_orientation}), we want to highlight in this section that on smaller scales, switchbacks can be deflected very consistently in the same direction. To that extent we report on a patch of switchbacks that occurs during encounter 2 from 2020-04-05T20:00 to 2020-04-06T12:00, with a total duration of 16~h, and displayed in Figure \ref{fig: patch}. In the top panel we display the radial and tangential component of the magnetic field $B_R$ and $B_T$, as well as the expected components of the Parker spiral magnetic field. The difference between model and data is lightly shaded. In the bottom panels, we plot the 2D distribution $f(B_R, B_T)$ and $f(B_R, B_N)$, with the color indicating the number of points inside each bin. An arrow indicates the average expected direction of the Parker spiral.
In these plots, it is clear that the magnetic field deviates in one direction during the entire patch, which is $B_T$ negative in the ecliptic plane. This corresponds to $+\phi$ direction with the notation adopted in this paper. The path taken to deflect and return to the Parker spiral remains unchanged within a given plane (here the ecliptic), rather than randomly in three dimensions.

This further confirms the results from \cite{Horbury_2020} who found that the larger switchbacks within a patch tend to deflect in the same direction. Here we do not have a notion of switchback duration but show that deviations are comprised in the ecliptic plane ($\theta<30^{\circ}$, not shown) and one-sided with regard to the Parker spiral (+$\phi$ direction). This event interestingly goes in the opposite direction compared to the systematic bias we found in section \ref{sec:3_orientation}. This is not unexpected, as the data displayed in Figures \ref{fig: 3_E2_distribution} to \ref{fig: 6_marginals} shows that switchbacks may deflect in any directions, despite the average having a tendency toward negative $\phi$.

\section{Discussion} \label{sec:5_discussion}

\subsection{Parker spiral}
\label{sec:5.1}
We have shown in section \ref{sec:3.1} that as PSP's distance to the Sun decreases, the magnetic field data of quiet solar wind intervals seems to deviate from the Parker spiral model predictions. This is mainly visible in the data from E8 and E9 when PSP was diving down to 16~$R_{\odot}$ at perihelion (while data above 30 $R_{\odot}$ show no obvious trend, consistently with \cite{Badman_2021}). Geometrically, this means that we are overestimating the algebraic value of the Parker spiral angle $\alpha_p$ and that the spiral is less tightly wounded than expected. The Parker spiral model computed in the present study is given by equation \ref{eq: spiral}, with $\omega = 2.9\times10^{-6}$~s$^{-1}$, $r_0= 10~R_{\odot}$ and where $V_r(t)$ is the measured radial speed of the solar wind processed with a 2h low pass filter. However, this model implicitly assumes a constant solar wind speed between the source surface of radius $r_0$ and the spacecraft, and this hypothesis is likely no longer valid so close to the Sun, especially in the slow solar wind that accelerates until 10-20 solar radii (see e.g. \cite{Bruno_97}). 
With these values in mind, and seeing that the average value of solar wind speed during E8 and E9 is around 200 to 300~km/s, it is highly probable that at such heights, PSP is located within the acceleration region of the solar wind (recently, \cite{Kasper_2021}  reported that PSP even went down to the magnetically dominated corona during the latest orbits). This is consistent with our results, since the spiral we observe is straighter than the expected Parker spiral associated with the wind speed measured by PSP. Indeed, overestimating the algebraic value of $\alpha_p$ amounts to overestimating the value of the solar wind speed $V_r$ from the source 

\subsection{Switchback orientation}
\label{sec:5.2}
\begin{figure}[ht]
    \centering
    \includegraphics[width=.4\textwidth]{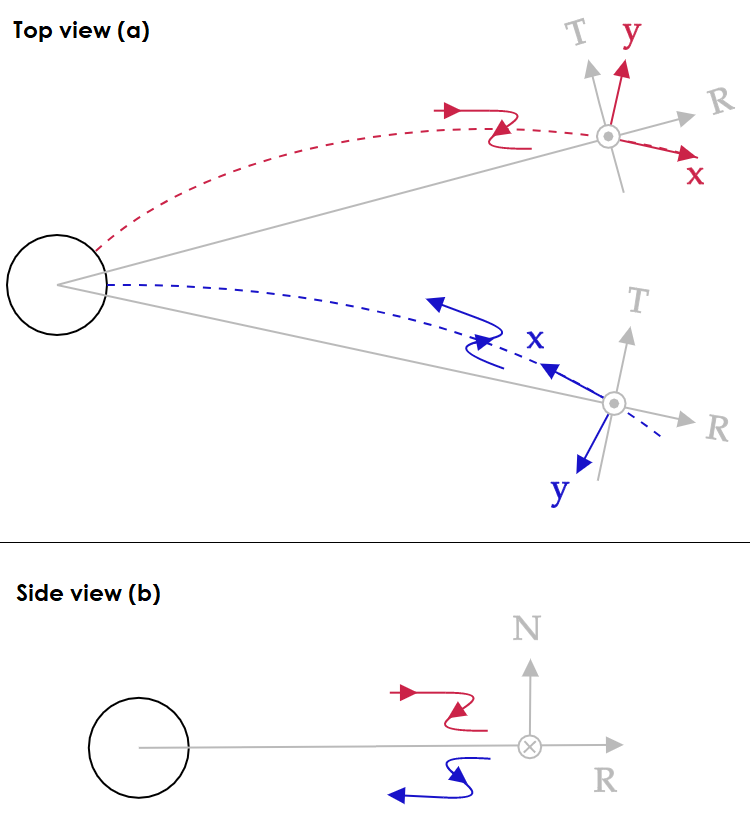}
    \caption{Illustration of the favored geometry of switchbacks in negative (blue) and positive (red) polarity}
    \label{fig: 7_orientation}
\end{figure}

In section \ref{sec:3_orientation}, we show that for all encounters (1 to 9, 3$^{\textrm{rd}}$ excluded), the switchback population presents a preferential deflection orientation toward lower values of the $\phi$ and $\theta$ angles. This result holds for all encounters (albeit being less clear in E6) and is not impacted by the polarity of the magnetic field.
We highlight the implication of this result in a more visual manner in Figure \ref{fig: 7_orientation}. In this sketch we represent in panel a) a top view of the Sun (N is in the out of plane direction), two field lines with positive (red) and negative (blue) polarity, and the associated Parker frame at a given radius as previously defined in section \ref{sec:2.3}. It is easier to see in this visualization that for a positive field (red), negative $\phi$ values correspond to the $\mathbf{-T}$ direction, while for a negative polarity field (blue) it corresponds to the $\mathbf{+T}$ direction (except for very large deflections close to $-\pi$ where the T component would reverse in both cases), both of these situations correspond to a clockwise rotation. This is consistent with the results of \cite{Horbury_2020} stating that switchbacks present a preferential orientation in the +T direction during encounter 1, where PSP samples mainly the negative polarity hemisphere of the Sun. This clockwise preference was observed in Helios data by \cite{MacNeil_2020}, and more recently identified by \cite{Meng_2022} in encounters 1 and 2 in PSP data, which further confirms our result. On Figure \ref{fig: 7_orientation} we draw a switchback illustration consistent with the negative $\phi$ preference for each polarity, and one can see that the geometry remains unchanged. In addition, switchbacks are both accelerated structures (sometimes called 'velocity spikes' due to their associated increase in $V_r$ \citep{Kasper_2019}) and Alfvénic properties. Hence, in the negative (respectively positive) magnetic sector, the magnetic field is correlated (anti-correlated) with the velocity vector. This leads to the field line configurations displayed in Figure \ref{fig: 7_orientation} are associated with positive tangential flows. On the other hand the less marked bias toward $-\theta$ values corresponds to the -N direction regardless of the polarity. One can realize that in this case, this indicates a symmetry of the switchback geometry in the two hemispheres. We illustrate this configuration in panel b, with a side view of the Sun (T is in the in-plane direction) and a switchback with negative $B_N$ for each polarity. To summarize, we find that switchbacks - viewed as a population of large magnetic deflections with respect to the Parker spiral - occur in all directions, while their deflection distribution presents a systematic bias in the $-\phi$ direction and to a lesser extent in the $-\theta$ direction. We now discuss this result in the light of the existing potential formation process for magnetic switchbacks presented in the introduction (cf section \ref{sec:1_intro}).

\begin{figure*}[ht]
    \centering
    \includegraphics[width=1\textwidth]{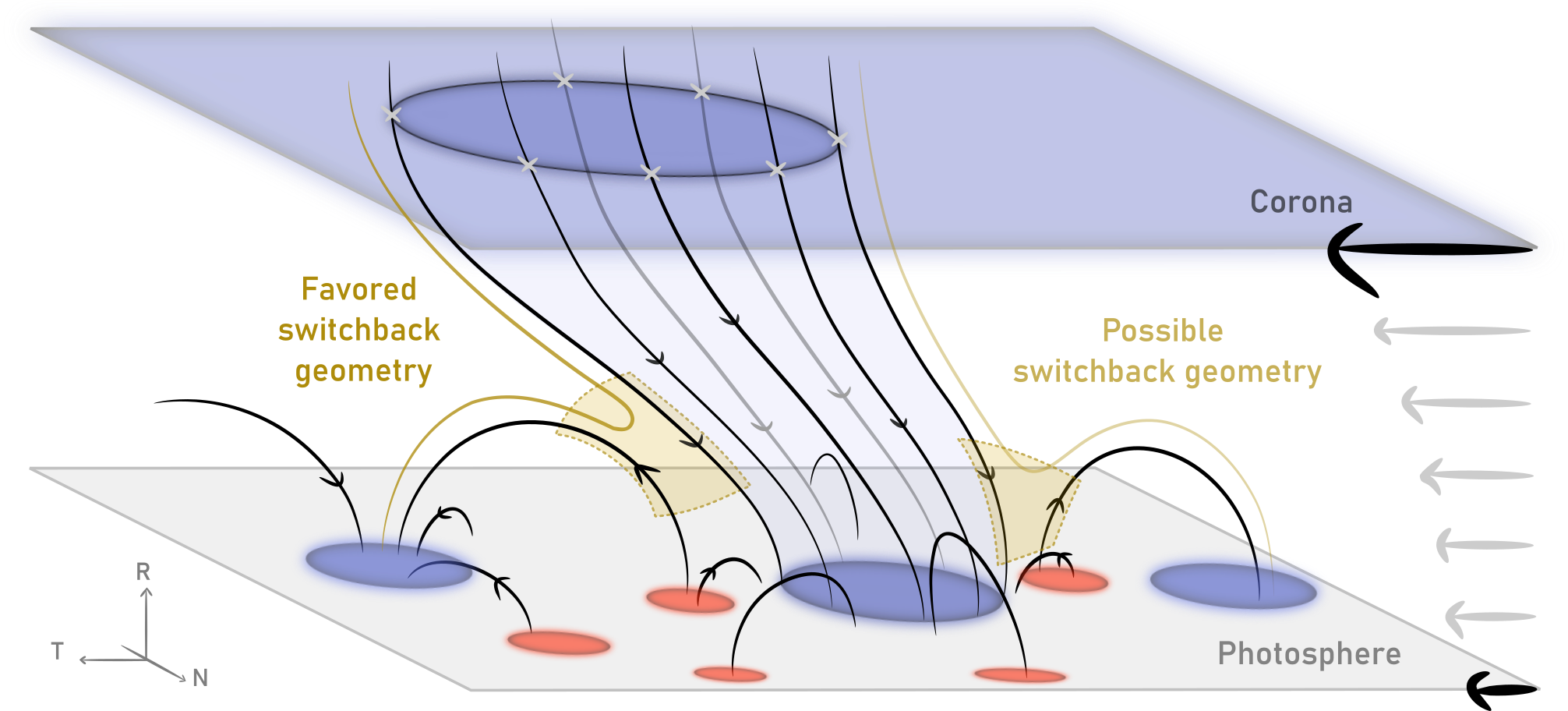}
    \caption{Illustration of a possible explanation for the preferential orientation of switchback, assuming interchange reconnection in the low corona as the initial mechanism. The sketch displays an element of open flux tubes from the photosphere expanding out into a faster corona, along closed loops forming the magnetic carpet. The blue (red) color is associated with the negative (positive) polarity of the field lines. Some potential reconnection sites are highlighted in light yellow (non exhaustive) and reconnected field lines are displayed in yellow as well. The arrows on the right stress the relative speed gradient that exists between the differentially rotating photosphere and the quasi-rigidly rotating corona.}
    \label{fig: 10_interpretation}
\end{figure*}

\subsection{Possible interpretation}

Interchange reconnection is a mechanism that allows the open magnetic field lines of the Sun to reconnect at their base with closed loops emerging from the magnetic carpet. This phenomenon mitigates the shear induced by the differential rotation of the photosphere - where field line footpoints rotate at different speeds depending on latitude - and the quasi-rigid rotation of the corona at equatorial rates - due to force balance with the large-scale coronal structure, including transients \citep{Wang_1996, Fisk_1996, Fisk_1999_APJ, Wang_and_sheeley_2004, Lionello_2005, Lionello_2006}. 
In the reference frame of a coronal hole that is corotating quasi-rigidly at the equatorial rotation rate, magnetic loops appear to drift in the direction opposite to that of solar rotation, from West to East. This relative drift can induce strong magnetic shears that force magnetic reconnection between magnetic loops and open field lines.
Subsequently, this leads to a footpoint displacement due to magnetic reconnection favored in the direction of solar rotation. Of course if the photosphere is going somehow faster than the corona (locally near the equator for instance) then the favored motion is reversed. In general, however, the process may be random and in all directions for the majority of the events because of localized photospheric motions associated with the magnetic carpet and solar granulation. The phenomenology at stake is illustrated in Figure \ref{fig: 10_interpretation}. There, the sketch displays an element of open flux tubes from the photosphere expanding out into a faster corona, inducing a shear in the magnetic field lines as just described. The sketch is valid for all such flux bundles that escape from the otherwise mixed polarity patchwork of closed field lines of the magnetic carpet. At the bottom of the flux tube, magnetic reconnection can occur randomly and in all directions between open field lines and closed loops emerging from the magnetic carpet. As the photosphere lags behind the solar corona, a particular geometry could be favored as footpoint motion will tend to mitigate the speed shear and jump in the direction of solar rotation. We suggest that this process could induce the bias in switchback orientation we present in this paper. This is consistent with \cite{Bale_2021} who also interpreted E6 data in terms of a shear between the photosphere and corona.

Our results here seem to be consistent with such reconnection occurring in regions where the photosphere is going on average slower than the solar corona and that would lead to the geometry highlighted in Figure \ref{fig: 7_orientation}a and \ref{fig: 10_interpretation}. This situation of a slower photosphere is particularly valid at mid to higher latitude, typically over 30$^{\circ}$ in latitude as studies of coronal hole rotation rate indicate (e.g.\cite{Giordano_2008, Mancuso_2011, Bagashvili_2017, Mancuso_2020}). However, analysis on the spacecraft connectivity throughout encounter 1 indicated that the measured solar wind observed by PSP was emerging from an equatorial coronal hole \citep{Bale_2019, Badman_2020, Reville_2020}, that would supposedly rotate close to the photospheric speed. We thus expect a lower $-\phi$ bias in this case. Nonetheless, we must consider the small but existing latitudinal extent of the coronal hole, as well as potential additional solar wind sources, in the interpretation of E1 data. Future work on the connectivity of PSP during switchback observation is needed for the different encounters, to confirm or infirm a potential link between the $-\phi$ bias and interchange reconnection induced by differential rotation.

Regarding the bias in elevation, \cite{Fisk_1999_APJ} interestingly highlight a potential circulation of field line footpoints at the photosphere from the poles toward the equator, which would be consistent with the slight bias we find toward negative $\theta$  values (i.e negative $B_N$, see Figure \ref{fig: 7_orientation}b). Indeed a field line rooted in the northern (southern) hemisphere would then be dragged downward (upward) and favor reconnection in the configuration displayed in red (blue) in Figure \ref{fig: 7_orientation}b. However, considering that the bias we find in $-\theta$ is small, we advise caution in the interpretation of this result and consider it less robust than the bias found in the ecliptic plane.

The preferential orientation we find - for switchbacks to deflect in the clockwise direction - does not seem to fit with a formation process involving either solely solar wind turbulence, as developed by \cite{Squire_2020, Mallet_2021, Shoda_2021}, or in-situ velocity shears as developed by \cite{Ruffolo_2020}. It seems that both of these processes would produce fluctuations that should appear as rather isotropic in the data. However, most of the studies cited above focus on the radial component of the magnetic field only. An analysis of the distribution of the magnetic field orientation angles in simulations from \cite{Squire_2020, Mallet_2021, Shoda_2021} (for turbulent generation) and \cite{Ruffolo_2020} (for in-situ velocity shears) would be of interest here, to investigate whether these other mechanisms can also introduce anisotropy in switchback properties.

We conclude that our results seem overall consistent with interchange reconnection in the low atmosphere as a plausible source of the preferential orientation of switchbacks. The bias we find is indeed going in the direction to reconcile the differential rotation of the photosphere and a more rigid rotation of the corona. 
Let us point out that we studied the switchback phenomenon in a probabilistic approach, without identifying exact structures in the data. Hence we can not conclude if the bias we find is due to switchbacks appearing more frequently in this direction, or if longer switchbacks tend to orient themselves in this direction. Finally, we realize that our study is not sufficient to determine how reconnection would create, propagate and preserve the switchbacks all the way to PSP's location (see \cite{Tenerani_2020} for instance). Several explanations stemming from interchange reconnection are currently investigated. We here provide an additional observational constraint, consistent with the results from \cite{Horbury_2020}, that models and simulations ought to reproduce. 

\section{Conclusion} \label{sec:5_conclusion}

We investigate a potential preferred orientation in the large magnetic deflections called switchbacks. 

We first caution that the choice of definition used to identify a magnetic switchback will by construction impact the results (section \ref{sec:2.2}). We choose to consider fluctuations away from the Parker spiral by using a locally defined Parker frame and two orientation angles in azimuth ($\phi$) and elevation ($\theta$) (section \ref{sec:2.3}).

We characterize the quiet solar wind orientation (section \ref{sec:3.1}) and find that the Parker spiral model indeed remains accurate at such short distances from the Sun. We notice that an offset appears for the latest encounters (8 and 9) and is linked to the lower radial distance. This is expected and shows that PSP is located near the acceleration region of the solar wind. 

We then investigate the large fluctuation orientation (section \ref{sec:3.2}). To do so, we assume that the wind is composed  of  two  populations  with  distinct  distribution  properties, respectively representing the background and perturbed solar winds. We assume a normal distribution of orientation angles for both distribution and fit our data with this model. This method allows us to define the switchback population without having to choose an arbitrary threshold in the magnetic field deviation. We find that the actual distribution of orientation angles is then well reproduced. We derive from this fit that the mean value of switchback population is biased by a few degrees toward lower $\phi$ for all encounters except E6 (-5.5$^{\circ}$ shift on average), and toward lower $\theta$ for all encounters but E2 and E6 (-2.1$^{\circ}$ shift on average, see Figure \ref{fig: 6_bias} and table \ref{tab:shift}). This occurs regardless of the main polarity of the field. We conclude that switchbacks occur in all direction, but present a preferential orientation in the $-\phi$ (clockwise) direction, and to a lesser extent in $-\theta$ (toward the equator) direction.

We report the observation of a patch of magnetic switchbacks who consistently deflected in the same direction over 16h. The deflections where all comprised with the ecliptic plane and on-sided regarding the Parker spiral. 

We discuss the implications of the preferred orientation we find (section \ref{sec:5_discussion}), showing that it favors an invariant geometry in the equatorial plane associated with a clockwise rotation and positive $V_t$ flows, while it may favor a symmetrical geometry  north and south of the HCS (Figure \ref{fig: 7_orientation}). These results are globally consistent with the observations of \cite{Horbury_2020, MacNeil_2020, Meng_2022}. The bias in $-\phi$ might find its cause in the interchange reconnection process occurring in the low corona, reconciling the shear induced by the different rotation rates of the photosphere and the corona.


\begin{acknowledgements}
    We acknowledge the NASA Parker Solar Probe Mission and particularly the FIELDS team led by S. D. Bale and the SWEAP team led by J. Kasper for use of data. Parker Solar Probe was designed, built, and is now operated by the Johns Hopkins Applied Physics Laboratory as part of NASA’s Living with a Star (LWS) program (contract NNN06AA01C). Support from the LWS management and technical team has played a critical role in the success of the Parker Solar Probe mission. The data used in this study are available at the NASA Space Physics Data Facility (SPDF): \href{https://spdf.gsfc.nasa.gov}{https://spdf.gsfc.nasa.gov}, and at the PSP science gateway \href{https://sppgway.jhuapl.edu/}{https://sppgway.jhuapl.edu/}. We visualize data using the CLWeb software (\href{ http://clweb.irap.omp.eu/}{ http://clweb.irap.omp.eu/}) developed by E. Penou; as well as the AMDA science analysis system (\href{http://amda.cdpp.eu/}{http://amda.cdpp.eu/}) provided by the Centre de Données de la Physique des Plasmas (CDPP) supported by CNRS, CNES, Observatoire de Paris and Université Paul Sabatier (UPS), Toulouse. Work at IRAP was supported by CNRS, CNES and UPS. The work of A. P. Rouillard and V. Réville was funded by the ERC SLOW\_SOURCE project (SLOW\_SOURCE—DLV-819189). The author N. Fargette acknowledges the support of the ISSI team, working in unraveling solar wind microphysics in the inner heliosphere. The author N. Fargette would finally like to aknowledge and thank P. Houdayer for helpful hindsight and discussions, as well as participation in the technical design of Figure 10. 
\end{acknowledgements}

\bibliographystyle{aa}
\bibliography{SOURCES}

\begin{thebibliography}{59}
\expandafter\ifx\csname natexlab\endcsname\relax\def\natexlab#1{#1}\fi

\bibitem[{{Akhavan-Tafti} {et~al.}(2021){Akhavan-Tafti}, {Kasper}, {Huang}, \&
  {Bale}}]{Akhavan-Tafti_2021}
{Akhavan-Tafti}, M., {Kasper}, J., {Huang}, J., \& {Bale}, S. 2021, \aap, 650,
  A4

\bibitem[{{Badman} {et~al.}(2020){Badman}, {Bale}, {Mart{\'\i}nez Oliveros},
  {Panasenco}, {Velli}, {Stansby}, {Buitrago-Casas}, {R{\'e}ville}, {Bonnell},
  {Case}, {Dudok de Wit}, {Goetz}, {Harvey}, {Kasper}, {Korreck}, {Larson},
  {Livi}, {MacDowall}, {Malaspina}, {Pulupa}, {Stevens}, \&
  {Whittlesey}}]{Badman_2020}
{Badman}, S.~T., {Bale}, S.~D., {Mart{\'\i}nez Oliveros}, J.~C., {et~al.} 2020,
  \apjs, 246, 23

\bibitem[{{Badman} {et~al.}(2021){Badman}, {Bale}, {Rouillard}, {Bowen},
  {Bonnell}, {Goetz}, {Harvey}, {MacDowall}, {Malaspina}, \&
  {Pulupa}}]{Badman_2021}
{Badman}, S.~T., {Bale}, S.~D., {Rouillard}, A.~P., {et~al.} 2021, \aap, 650,
  A18

\bibitem[{{Bagashvili} {et~al.}(2017){Bagashvili}, {Shergelashvili},
  {Japaridze}, {Chargeishvili}, {Kosovichev}, {Kukhianidze}, {Ramishvili},
  {Zaqarashvili}, {Poedts}, {Khodachenko}, \& {De
  Causmaecker}}]{Bagashvili_2017}
{Bagashvili}, S.~R., {Shergelashvili}, B.~M., {Japaridze}, D.~R., {et~al.}
  2017, \aap, 603, A134

\bibitem[{{Bale} {et~al.}(2019){Bale}, {Badman}, {Bonnell}, {Bowen}, {Burgess},
  {Case}, {Cattell}, {Chandran}, {Chaston}, {Chen}, {Drake}, {de Wit},
  {Eastwood}, {Ergun}, {Farrell}, {Fong}, {Goetz}, {Goldstein}, {Goodrich},
  {Harvey}, {Horbury}, {Howes}, {Kasper}, {Kellogg}, {Klimchuk}, {Korreck},
  {Krasnoselskikh}, {Krucker}, {Laker}, {Larson}, {MacDowall}, {Maksimovic},
  {Malaspina}, {Martinez-Oliveros}, {McComas}, {Meyer-Vernet}, {Moncuquet},
  {Mozer}, {Phan}, {Pulupa}, {Raouafi}, {Salem}, {Stansby}, {Stevens}, {Szabo},
  {Velli}, {Woolley}, \& {Wygant}}]{Bale_2019}
{Bale}, S.~D., {Badman}, S.~T., {Bonnell}, J.~W., {et~al.} 2019, \nat, 576, 237

\bibitem[{{Bale} {et~al.}(2016){Bale}, {Goetz}, {Harvey}, {Turin}, {Bonnell},
  {Dudok de Wit}, {Ergun}, {MacDowall}, {Pulupa}, {Andre}, {Bolton},
  {Bougeret}, {Bowen}, {Burgess}, {Cattell}, {Chandran}, {Chaston}, {Chen},
  {Choi}, {Connerney}, {Cranmer}, {Diaz-Aguado}, {Donakowski}, {Drake},
  {Farrell}, {Fergeau}, {Fermin}, {Fischer}, {Fox}, {Glaser}, {Goldstein},
  {Gordon}, {Hanson}, {Harris}, {Hayes}, {Hinze}, {Hollweg}, {Horbury},
  {Howard}, {Hoxie}, {Jannet}, {Karlsson}, {Kasper}, {Kellogg}, {Kien},
  {Klimchuk}, {Krasnoselskikh}, {Krucker}, {Lynch}, {Maksimovic}, {Malaspina},
  {Marker}, {Martin}, {Martinez-Oliveros}, {McCauley}, {McComas}, {McDonald},
  {Meyer-Vernet}, {Moncuquet}, {Monson}, {Mozer}, {Murphy}, {Odom},
  {Oliverson}, {Olson}, {Parker}, {Pankow}, {Phan}, {Quataert}, {Quinn},
  {Ruplin}, {Salem}, {Seitz}, {Sheppard}, {Siy}, {Stevens}, {Summers}, {Szabo},
  {Timofeeva}, {Vaivads}, {Velli}, {Yehle}, {Werthimer}, \&
  {Wygant}}]{Bale_2016}
{Bale}, S.~D., {Goetz}, K., {Harvey}, P.~R., {et~al.} 2016, \ssr, 204, 49

\bibitem[{{Bale} {et~al.}(2021){Bale}, {Horbury}, {Velli}, {Desai}, {Halekas},
  {McManus}, {Panasenco}, {Badman}, {Bowen}, {Chandran}, {Drake}, {Kasper},
  {Laker}, {Mallet}, {Matteini}, {Phan}, {Raouafi}, {Squire}, {Woodham}, \&
  {Woolley}}]{Bale_2021}
{Bale}, S.~D., {Horbury}, T.~S., {Velli}, M., {et~al.} 2021, \apj, 923, 174

\bibitem[{{Balogh} {et~al.}(1999){Balogh}, {Forsyth}, {Lucek}, {Horbury}, \&
  {Smith}}]{Balogh_1999}
{Balogh}, A., {Forsyth}, R.~J., {Lucek}, E.~A., {Horbury}, T.~S., \& {Smith},
  E.~J. 1999, \grl, 26, 631

\bibitem[{{Bandyopadhyay} {et~al.}(2021){Bandyopadhyay}, {Matthaeus},
  {McComas}, {Joyce}, {Szalay}, {Christian}, {Giacalone}, {Schwadron},
  {Mitchell}, {Hill}, {McNutt}, {Desai}, {Bale}, {Bonnell}, {Dudok de Wit},
  {Goetz}, {Harvey}, {MacDowall}, {Malaspina}, {Pulupa}, {Kasper}, \&
  {Stevens}}]{Bandyopadhyay_2021}
{Bandyopadhyay}, R., {Matthaeus}, W.~H., {McComas}, D.~J., {et~al.} 2021, \aap,
  650, L4

\bibitem[{{Bourouaine} {et~al.}(2020){Bourouaine}, {Perez}, {Klein}, {Chen},
  {Martinovi{\'c}}, {Bale}, {Kasper}, \& {Raouafi}}]{Bourouaine_2020}
{Bourouaine}, S., {Perez}, J.~C., {Klein}, K.~G., {et~al.} 2020, \apjl, 904,
  L30

\bibitem[{{Bruno} \& {Bavassano}(1997)}]{Bruno_97}
{Bruno}, R. \& {Bavassano}, B. 1997, \grl, 24, 2267

\bibitem[{{Case} {et~al.}(2020){Case}, {Kasper}, {Stevens}, {Korreck},
  {Paulson}, {Daigneau}, {Caldwell}, {Freeman}, {Henry}, {Klingensmith},
  {Bookbinder}, {Robinson}, {Berg}, {Tiu}, {Wright}, {Reinhart}, {Curtis},
  {Ludlam}, {Larson}, {Whittlesey}, {Livi}, {Klein}, \&
  {Martinovi{\'c}}}]{2020ApJS..246...43C}
{Case}, A.~W., {Kasper}, J.~C., {Stevens}, M.~L., {et~al.} 2020, \apjs, 246, 43

\bibitem[{{Drake} {et~al.}(2021){Drake}, {Agapitov}, {Swisdak}, {Badman},
  {Bale}, {Horbury}, {Kasper}, {MacDowall}, {Mozer}, {Phan}, {Pulupa}, {Szabo},
  \& {Velli}}]{Drake_2021}
{Drake}, J.~F., {Agapitov}, O., {Swisdak}, M., {et~al.} 2021, \aap, 650, A2

\bibitem[{{Dudok de Wit} {et~al.}(2020){Dudok de Wit}, {Krasnoselskikh},
  {Bale}, {Bonnell}, {Bowen}, {Chen}, {Froment}, {Goetz}, {Harvey},
  {Jagarlamudi}, {Larosa}, {MacDowall}, {Malaspina}, {Matthaeus}, {Pulupa},
  {Velli}, \& {Whittlesey}}]{DDW_2020}
{Dudok de Wit}, T., {Krasnoselskikh}, V.~V., {Bale}, S.~D., {et~al.} 2020,
  \apjs, 246, 39

\bibitem[{{Fargette} {et~al.}(2021){Fargette}, {Lavraud}, {Rouillard},
  {R{\'e}ville}, {Dudok De Wit}, {Froment}, {Halekas}, {Phan}, {Malaspina},
  {Bale}, {Kasper}, {Louarn}, {Case}, {Korreck}, {Larson}, {Pulupa}, {Stevens},
  {Whittlesey}, \& {Berthomier}}]{Fargette_2021}
{Fargette}, N., {Lavraud}, B., {Rouillard}, A.~P., {et~al.} 2021, \apj, 919, 96

\bibitem[{{Fisk}(1996)}]{Fisk_1996}
{Fisk}, L.~A. 1996, \jgr, 101, 15547

\bibitem[{{Fisk} \& {Kasper}(2020)}]{Fisk_and_Kasper_2020}
{Fisk}, L.~A. \& {Kasper}, J.~C. 2020, \apjl, 894, L4

\bibitem[{{Fisk} {et~al.}(1999){Fisk}, {Zurbuchen}, \&
  {Schwadron}}]{Fisk_1999_APJ}
{Fisk}, L.~A., {Zurbuchen}, T.~H., \& {Schwadron}, N.~A. 1999, \apj, 521, 868

\bibitem[{{Foreman-Mackey} {et~al.}(2019){Foreman-Mackey}, {Farr}, {Sinha},
  {Archibald}, {Hogg}, {Sanders}, {Zuntz}, {Williams}, {Nelson}, {de
  Val-Borro}, {Erhardt}, {Pashchenko}, \& {Pla}}]{Foreman-Mackey_2019}
{Foreman-Mackey}, D., {Farr}, W., {Sinha}, M., {et~al.} 2019, The Journal of
  Open Source Software, 4, 1864

\bibitem[{{Giordano} \& {Mancuso}(2008)}]{Giordano_2008}
{Giordano}, S. \& {Mancuso}, S. 2008, \apj, 688, 656

\bibitem[{{Gosling} {et~al.}(2011){Gosling}, {Tian}, \& {Phan}}]{Gosling_2011}
{Gosling}, J.~T., {Tian}, H., \& {Phan}, T.~D. 2011, \apjl, 737, L35

\bibitem[{{Horbury} {et~al.}(2018){Horbury}, {Matteini}, \&
  {Stansby}}]{Horbury_2018}
{Horbury}, T.~S., {Matteini}, L., \& {Stansby}, D. 2018, \mnras, 478, 1980

\bibitem[{{Horbury} {et~al.}(2020){Horbury}, {Woolley}, {Laker}, {Matteini},
  {Eastwood}, {Bale}, {Velli}, {Chandran}, {Phan}, {Raouafi}, {Goetz},
  {Harvey}, {Pulupa}, {Klein}, {Dudok de Wit}, {Kasper}, {Korreck}, {Case},
  {Stevens}, {Whittlesey}, {Larson}, {MacDowall}, {Malaspina}, \&
  {Livi}}]{Horbury_2020}
{Horbury}, T.~S., {Woolley}, T., {Laker}, R., {et~al.} 2020, \apjs, 246, 45

\bibitem[{{Kasper} {et~al.}(2016){Kasper}, {Abiad}, {Austin}, {Balat-Pichelin},
  {Bale}, {Belcher}, {Berg}, {Bergner}, {Berthomier}, {Bookbinder}, {Brodu},
  {Caldwell}, {Case}, {Chand ran}, {Cheimets}, {Cirtain}, {Cranmer}, {Curtis},
  {Daigneau}, {Dalton}, {Dasgupta}, {DeTomaso}, {Diaz-Aguado}, {Djordjevic},
  {Donaskowski}, {Effinger}, {Florinski}, {Fox}, {Freeman}, {Gallagher},
  {Gary}, {Gauron}, {Gates}, {Goldstein}, {Golub}, {Gordon}, {Gurnee}, {Guth},
  {Halekas}, {Hatch}, {Heerikuisen}, {Ho}, {Hu}, {Johnson}, {Jordan},
  {Korreck}, {Larson}, {Lazarus}, {Li}, {Livi}, {Ludlam}, {Maksimovic},
  {McFadden}, {Marchant}, {Maruca}, {McComas}, {Messina}, {Mercer}, {Park},
  {Peddie}, {Pogorelov}, {Reinhart}, {Richardson}, {Robinson}, {Rosen},
  {Skoug}, {Slagle}, {Steinberg}, {Stevens}, {Szabo}, {Taylor}, {Tiu}, {Turin},
  {Velli}, {Webb}, {Whittlesey}, {Wright}, {Wu}, \&
  {Zank}}]{2016SSRv..204..131K}
{Kasper}, J.~C., {Abiad}, R., {Austin}, G., {et~al.} 2016, \ssr, 204, 131

\bibitem[{{Kasper} {et~al.}(2019){Kasper}, {Bale}, {Belcher}, {Berthomier},
  {Case}, {Chandran}, {Curtis}, {Gallagher}, {Gary}, {Golub}, {Halekas}, {Ho},
  {Horbury}, {Hu}, {Huang}, {Klein}, {Korreck}, {Larson}, {Livi}, {Maruca},
  {Lavraud}, {Louarn}, {Maksimovic}, {Martinovic}, {McGinnis}, {Pogorelov},
  {Richardson}, {Skoug}, {Steinberg}, {Stevens}, {Szabo}, {Velli},
  {Whittlesey}, {Wright}, {Zank}, {MacDowall}, {McComas}, {McNutt}, {Pulupa},
  {Raouafi}, \& {Schwadron}}]{Kasper_2019}
{Kasper}, J.~C., {Bale}, S.~D., {Belcher}, J.~W., {et~al.} 2019, \nat, 576, 228

\bibitem[{{Kasper} {et~al.}(2021){Kasper}, {Klein}, {Lichko}, {Huang}, {Chen},
  {Badman}, {Bonnell}, {Whittlesey}, {Livi}, {Larson}, {Pulupa}, {Rahmati},
  {Stansby}, {Korreck}, {Stevens}, {Case}, {Bale}, {Maksimovic}, {Moncuquet},
  {Goetz}, {Halekas}, {Malaspina}, {Raouafi}, {Szabo}, {MacDowall}, {Velli},
  {Dudok de Wit}, \& {Zank}}]{Kasper_2021}
{Kasper}, J.~C., {Klein}, K.~G., {Lichko}, E., {et~al.} 2021, \prl, 127, 255101

\bibitem[{{Laker} {et~al.}(2021){Laker}, {Horbury}, {Bale}, {Matteini},
  {Woolley}, {Woodham}, {Badman}, {Pulupa}, {Kasper}, {Stevens}, {Case}, \&
  {Korreck}}]{Laker_2021}
{Laker}, R., {Horbury}, T.~S., {Bale}, S.~D., {et~al.} 2021, \aap, 650, A1

\bibitem[{{Larosa} {et~al.}(2021){Larosa}, {Krasnoselskikh}, {Dudok de Wit},
  {Agapitov}, {Froment}, {Jagarlamudi}, {Velli}, {Bale}, {Case}, {Goetz},
  {Harvey}, {Kasper}, {Korreck}, {Larson}, {MacDowall}, {Malaspina}, {Pulupa},
  {Revillet}, \& {Stevens}}]{Larosa_2021}
{Larosa}, A., {Krasnoselskikh}, V., {Dudok de Wit}, T., {et~al.} 2021, \aap,
  650, A3

\bibitem[{{Lionello} {et~al.}(2006){Lionello}, {Linker}, {Miki{\'c}}, \&
  {Riley}}]{Lionello_2006}
{Lionello}, R., {Linker}, J.~A., {Miki{\'c}}, Z., \& {Riley}, P. 2006, \apjl,
  642, L69

\bibitem[{{Lionello} {et~al.}(2005){Lionello}, {Riley}, {Linker}, \&
  {Miki{\'c}}}]{Lionello_2005}
{Lionello}, R., {Riley}, P., {Linker}, J.~A., \& {Miki{\'c}}, Z. 2005, \apj,
  625, 463

\bibitem[{Livi {et~al.}(2021)Livi, Larson, Kasper, Abiad, Case, Klein, Curtis,
  Dalton, Stevens, Korreck, \& et~al.}]{Livi_2021}
Livi, R., Larson, D.~E., Kasper, J.~C., {et~al.} 2021, Earth and Space Science
  Open Archive, 20

\bibitem[{{Macneil} {et~al.}(2020){Macneil}, {Owens}, {Wicks}, {Lockwood},
  {Bentley}, \& {Lang}}]{MacNeil_2020}
{Macneil}, A.~R., {Owens}, M.~J., {Wicks}, R.~T., {et~al.} 2020, \mnras, 494,
  3642

\bibitem[{{Mallet} {et~al.}(2021){Mallet}, {Squire}, {Chandran}, {Bowen}, \&
  {Bale}}]{Mallet_2021}
{Mallet}, A., {Squire}, J., {Chandran}, B. D.~G., {Bowen}, T., \& {Bale}, S.~D.
  2021, \apj, 918, 62

\bibitem[{{Mancuso} \& {Giordano}(2011)}]{Mancuso_2011}
{Mancuso}, S. \& {Giordano}, S. 2011, \apj, 729, 79

\bibitem[{{Mancuso} {et~al.}(2020){Mancuso}, {Giordano}, {Barghini}, \&
  {Telloni}}]{Mancuso_2020}
{Mancuso}, S., {Giordano}, S., {Barghini}, D., \& {Telloni}, D. 2020, \aap,
  644, A18

\bibitem[{{Martinovi{\'c}} {et~al.}(2021){Martinovi{\'c}}, {Klein}, {Huang},
  {Chandran}, {Kasper}, {Lichko}, {Bowen}, {Chen}, {Matteini}, {Stevens},
  {Case}, \& {Bale}}]{Martinovic_2021}
{Martinovi{\'c}}, M.~M., {Klein}, K.~G., {Huang}, J., {et~al.} 2021, \apj, 912,
  28

\bibitem[{{Matteini} {et~al.}(2014){Matteini}, {Horbury}, {Neugebauer}, \&
  {Goldstein}}]{Matteini_2014}
{Matteini}, L., {Horbury}, T.~S., {Neugebauer}, M., \& {Goldstein}, B.~E. 2014,
  \grl, 41, 259

\bibitem[{{Meng} {et~al.}(2022){Meng}, {Liu}, {Chen}, \& {Wang}}]{Meng_2022}
{Meng}, M.-M., {Liu}, Y.~D., {Chen}, C., \& {Wang}, R. 2022, Research in
  Astronomy and Astrophysics, 22, 035018

\bibitem[{{Mozer} {et~al.}(2020){Mozer}, {Agapitov}, {Bale}, {Bonnell}, {Case},
  {Chaston}, {Curtis}, {Dudok de Wit}, {Goetz}, {Goodrich}, {Harvey}, {Kasper},
  {Korreck}, {Krasnoselskikh}, {Larson}, {Livi}, {MacDowall}, {Malaspina},
  {Pulupa}, {Stevens}, {Whittlesey}, \& {Wygant}}]{Mozer_2020}
{Mozer}, F.~S., {Agapitov}, O.~V., {Bale}, S.~D., {et~al.} 2020, \apjs, 246, 68

\bibitem[{{Nash} {et~al.}(1988){Nash}, {Sheeley}, \& {Wang}}]{Nash_1988}
{Nash}, A.~G., {Sheeley}, N.~R., J., \& {Wang}, Y.~M. 1988, \solphys, 117, 359

\bibitem[{{Owens} {et~al.}(2020){Owens}, {Lockwood}, {Macneil}, \&
  {Stansby}}]{Owens_2020}
{Owens}, M., {Lockwood}, M., {Macneil}, A., \& {Stansby}, D. 2020, \solphys,
  295, 37

\bibitem[{{Owens} {et~al.}(2018){Owens}, {Lockwood}, {Barnard}, \&
  {MacNeil}}]{Owens_2018}
{Owens}, M.~J., {Lockwood}, M., {Barnard}, L.~A., \& {MacNeil}, A.~R. 2018,
  \apjl, 868, L14

\bibitem[{{Parker}(1958)}]{Parker_58}
{Parker}, E.~N. 1958, \apj, 128, 664

\bibitem[{{Phan} {et~al.}(2020){Phan}, {Bale}, {Eastwood}, {Lavraud}, {Drake},
  {Oieroset}, {Shay}, {Pulupa}, {Stevens}, {MacDowall}, {Case}, {Larson},
  {Kasper}, {Whittlesey}, {Szabo}, {Korreck}, {Bonnell}, {de Wit}, {Goetz},
  {Harvey}, {Horbury}, {Livi}, {Malaspina}, {Paulson}, {Raouafi}, \&
  {Velli}}]{Phan_2020}
{Phan}, T.~D., {Bale}, S.~D., {Eastwood}, J.~P., {et~al.} 2020, \apjs, 246, 34

\bibitem[{{R{\'e}ville} {et~al.}(2020){R{\'e}ville}, {Velli}, {Panasenco},
  {Tenerani}, {Shi}, {Badman}, {Bale}, {Kasper}, {Stevens}, {Korreck},
  {Bonnell}, {Case}, {de Wit}, {Goetz}, {Harvey}, {Larson}, {Livi},
  {Malaspina}, {MacDowall}, {Pulupa}, \& {Whittlesey}}]{Reville_2020}
{R{\'e}ville}, V., {Velli}, M., {Panasenco}, O., {et~al.} 2020, \apjs, 246, 24

\bibitem[{{Ruffolo} {et~al.}(2020){Ruffolo}, {Matthaeus}, {Chhiber}, {Usmanov},
  {Yang}, {Bandyopadhyay}, {Parashar}, {Goldstein}, {DeForest}, {Wan},
  {Chasapis}, {Maruca}, {Velli}, \& {Kasper}}]{Ruffolo_2020}
{Ruffolo}, D., {Matthaeus}, W.~H., {Chhiber}, R., {et~al.} 2020, \apj, 902, 94

\bibitem[{{Schwadron} \& {McComas}(2021)}]{Schwadron_2021}
{Schwadron}, N.~A. \& {McComas}, D.~J. 2021, \apj, 909, 95

\bibitem[{{Shoda} {et~al.}(2021){Shoda}, {Chandran}, \& {Cranmer}}]{Shoda_2021}
{Shoda}, M., {Chandran}, B. D.~G., \& {Cranmer}, S.~R. 2021, \apj, 915, 52

\bibitem[{{Squire} {et~al.}(2020){Squire}, {Chandran}, \&
  {Meyrand}}]{Squire_2020}
{Squire}, J., {Chandran}, B.~D.~G., \& {Meyrand}, R. 2020, \apjl, 891, L2

\bibitem[{{Sterling} \& {Moore}(2020)}]{Sterling_2020}
{Sterling}, A.~C. \& {Moore}, R.~L. 2020, \apjl, 896, L18

\bibitem[{{Tenerani} {et~al.}(2020){Tenerani}, {Velli}, {Matteini},
  {R{\'e}ville}, {Shi}, {Bale}, {Kasper}, {Bonnell}, {Case}, {de Wit}, {Goetz},
  {Harvey}, {Klein}, {Korreck}, {Larson}, {Livi}, {MacDowall}, {Malaspina},
  {Pulupa}, {Stevens}, \& {Whittlesey}}]{Tenerani_2020}
{Tenerani}, A., {Velli}, M., {Matteini}, L., {et~al.} 2020, \apjs, 246, 32

\bibitem[{{Wang} {et~al.}(1996){Wang}, {Hawley}, \& {Sheeley}}]{Wang_1996}
{Wang}, Y.-M., {Hawley}, S.~H., \& {Sheeley}, Neil~R., J. 1996, Science, 271,
  464

\bibitem[{{Wang} {et~al.}(1989){Wang}, {Nash}, \& {Sheeley}}]{Wang_1989}
{Wang}, Y.~M., {Nash}, A.~G., \& {Sheeley}, N.~R., J. 1989, Science, 245, 712

\bibitem[{{Wang} \& {Sheeley}(2004)}]{Wang_and_sheeley_2004}
{Wang}, Y.~M. \& {Sheeley}, N.~R., J. 2004, \apj, 612, 1196

\bibitem[{{Whittlesey} {et~al.}(2020){Whittlesey}, {Larson}, {Kasper},
  {Halekas}, {Abatcha}, {Abiad}, {Berthomier}, {Case}, {Chen}, {Curtis},
  {Dalton}, {Klein}, {Korreck}, {Livi}, {Ludlam}, {Marckwordt}, {Rahmati},
  {Robinson}, {Slagle}, {Stevens}, {Tiu}, \& {Verniero}}]{2020ApJS..246...74W}
{Whittlesey}, P.~L., {Larson}, D.~E., {Kasper}, J.~C., {et~al.} 2020, \apjs,
  246, 74

\bibitem[{{Woodham} {et~al.}(2021){Woodham}, {Horbury}, {Matteini}, {Woolley},
  {Laker}, {Bale}, {Nicolaou}, {Stawarz}, {Stansby}, {Hietala}, {Larson},
  {Livi}, {Verniero}, {McManus}, {Kasper}, {Korreck}, {Raouafi}, {Moncuquet},
  \& {Pulupa}}]{Woodham_2021}
{Woodham}, L.~D., {Horbury}, T.~S., {Matteini}, L., {et~al.} 2021, \aap, 650,
  L1

\bibitem[{{Woolley} {et~al.}(2020){Woolley}, {Matteini}, {Horbury}, {Bale},
  {Woodham}, {Laker}, {Alterman}, {Bonnell}, {Case}, {Kasper}, {Klein},
  {Martinovi{\'c}}, \& {Stevens}}]{Woolley_2020}
{Woolley}, T., {Matteini}, L., {Horbury}, T.~S., {et~al.} 2020, \mnras, 498,
  5524

\bibitem[{{Wu} {et~al.}(2021){Wu}, {Tu}, {Wang}, \& {Yang}}]{Wu_2021}
{Wu}, H., {Tu}, C., {Wang}, X., \& {Yang}, L. 2021, \apj, 911, 73

\bibitem[{{Zank} {et~al.}(2020){Zank}, {Nakanotani}, {Zhao}, {Adhikari}, \&
  {Kasper}}]{Zank_2020}
{Zank}, G.~P., {Nakanotani}, M., {Zhao}, L.~L., {Adhikari}, L., \& {Kasper}, J.
  2020, \apj, 903, 1

\end{thebibliography}

\begin{appendix} 

\section{Quiet wind intervals}
\label{sec: app_A}
\bottomcaption{Timetable of quiet solar wind intervals manually selected over encounters 1, 2, 4, 5, 6, 7, 8 and 9 (see section \ref{sec:3.1})}

\begin{supertabular}{ccc}
\hline \hline
Enc & Start time & End time \\
\hline \hline
1 & 2018-10-31T23:17:00 & 2018-11-01T01:28:00 \\
1 & 2018-11-01T14:00:00 & 2018-11-01T17:00:00 \\
1 & 2018-11-02T08:00:00 & 2018-11-02T12:00:00 \\
1 & 2018-11-02T13:18:00 & 2018-11-02T17:00:00 \\
1 & 2018-11-02T20:37:00 & 2018-11-02T23:25:00 \\
1 & 2018-11-03T02:40:00 & 2018-11-03T05:15:00 \\
1 & 2018-11-03T14:58:00 & 2018-11-03T20:52:00 \\
1 & 2018-11-04T17:40:00 & 2018-11-05T01:23:00 \\
1 & 2018-11-05T18:00:00 & 2018-11-05T23:00:00 \\
1 & 2018-11-07T03:59:00 & 2018-11-07T05:56:00 \\
1 & 2018-11-08T07:04:00 & 2018-11-08T09:10:00 \\
1 & 2018-11-08T10:04:00 & 2018-11-08T12:07:00 \\
1 & 2018-11-08T17:46:00 & 2018-11-08T20:47:00 \\
1 & 2018-11-10T10:25:00 & 2018-11-10T18:31:00 \\
1 & 2018-11-11T00:00:00 & 2018-11-11T17:00:00 \\
\hline
2 & 2019-03-30T18:00:00 & 2019-03-30T19:18:00 \\
2 & 2019-03-30T21:10:00 & 2019-03-30T22:00:00 \\
2 & 2019-04-01T06:00:00 & 2019-04-01T09:00:00 \\
2 & 2019-04-01T18:16:00 & 2019-04-01T21:47:00 \\
2 & 2019-04-02T06:46:00 & 2019-04-02T08:14:00 \\
2 & 2019-04-02T10:34:00 & 2019-04-02T11:43:00 \\
2 & 2019-04-02T14:04:00 & 2019-04-02T15:06:00 \\
2 & 2019-04-02T18:07:47 & 2019-04-02T18:59:53 \\
2 & 2019-04-03T02:31:00 & 2019-04-03T03:23:00 \\
2 & 2019-04-03T08:00:00 & 2019-04-03T11:00:00 \\
2 & 2019-04-03T13:00:00 & 2019-04-03T20:00:00 \\
2 & 2019-04-03T23:02:00 & 2019-04-04T01:43:00 \\
2 & 2019-04-04T04:43:00 & 2019-04-04T06:50:00 \\
2 & 2019-04-04T15:06:00 & 2019-04-04T16:22:00 \\
2 & 2019-04-04T17:59:00 & 2019-04-04T18:55:00 \\
2 & 2019-04-05T18:46:00 & 2019-04-06T00:32:00 \\
2 & 2019-04-06T10:00:00 & 2019-04-06T13:00:00 \\
2 & 2019-04-07T10:00:00 & 2019-04-07T19:00:00 \\
2 & 2019-04-08T18:09:00 & 2019-04-08T20:18:00 \\
2 & 2019-04-09T19:00:00 & 2019-04-09T20:29:00 \\
2 & 2019-04-10T14:59:00 & 2019-04-10T17:31:00 \\
2 & 2019-04-11T09:00:00 & 2019-04-11T12:00:00 \\
\hline 
4 & 2020-01-22T15:00:00 & 2020-01-22T16:00:00 \\
4 & 2020-01-23T15:00:00 & 2020-01-23T16:00:00 \\
4 & 2020-01-23T17:05:00 & 2020-01-23T18:38:00 \\
4 & 2020-01-23T23:04:00 & 2020-01-24T00:53:00 \\
4 & 2020-01-24T05:00:00 & 2020-01-24T06:00:00 \\
4 & 2020-01-24T16:00:00 & 2020-01-24T19:00:00 \\
4 & 2020-01-25T06:01:21 & 2020-01-25T06:55:26 \\
4 & 2020-01-25T23:00:00 & 2020-01-26T00:32:00 \\
4 & 2020-01-26T05:04:00 & 2020-01-26T06:58:00 \\
4 & 2020-01-26T14:00:00 & 2020-01-26T15:00:00 \\
4 & 2020-01-27T05:19:00 & 2020-01-27T06:41:00 \\
4 & 2020-01-28T05:00:00 & 2020-01-28T06:00:00 \\
4 & 2020-01-28T10:00:00 & 2020-01-28T12:00:00 \\
4 & 2020-01-28T23:59:00 & 2020-01-29T01:35:00 \\
4 & 2020-01-29T01:44:00 & 2020-01-29T03:00:00 \\
4 & 2020-01-29T11:00:00 & 2020-01-29T13:00:00 \\
4 & 2020-01-29T16:00:00 & 2020-01-29T23:00:00 \\
4 & 2020-01-30T05:39:00 & 2020-01-30T06:56:00 \\
4 & 2020-01-30T07:04:00 & 2020-01-30T08:50:00 \\
4 & 2020-01-30T09:09:00 & 2020-01-30T11:00:00 \\
4 & 2020-01-30T12:00:00 & 2020-01-30T13:00:00 \\
4 & 2020-01-30T22:00:00 & 2020-01-31T01:00:00 \\
4 & 2020-01-31T12:17:00 & 2020-01-31T17:59:00 \\
4 & 2020-02-01T20:11:43 & 2020-02-01T20:59:39 \\
4 & 2020-02-02T00:00:00 & 2020-02-02T01:42:00 \\
4 & 2020-02-02T10:00:00 & 2020-02-02T12:00:00 \\
4 & 2020-02-03T14:00:00 & 2020-02-03T17:00:00 \\
\hline
5 & 2020-06-01T07:00:00 & 2020-06-01T09:00:00 \\
5 & 2020-06-01T16:53:00 & 2020-06-01T18:10:00 \\
5 & 2020-06-02T00:10:00 & 2020-06-02T04:54:00 \\
5 & 2020-06-02T20:14:00 & 2020-06-03T00:04:00 \\
5 & 2020-06-03T16:13:00 & 2020-06-03T17:51:00 \\
5 & 2020-06-04T00:55:00 & 2020-06-04T03:17:00 \\
5 & 2020-06-04T06:21:00 & 2020-06-04T07:32:00 \\
5 & 2020-06-04T10:19:00 & 2020-06-04T12:12:00 \\
5 & 2020-06-04T18:04:00 & 2020-06-04T19:43:00 \\
5 & 2020-06-06T09:41:00 & 2020-06-06T12:25:00 \\
5 & 2020-06-07T00:27:00 & 2020-06-07T02:01:00 \\
5 & 2020-06-07T04:19:00 & 2020-06-07T06:14:00 \\
5 & 2020-06-07T10:00:00 & 2020-06-07T11:00:00 \\
5 & 2020-06-07T13:00:00 & 2020-06-07T15:00:00 \\
5 & 2020-06-07T16:00:00 & 2020-06-07T19:00:00 \\
5 & 2020-06-08T12:48:00 & 2020-06-08T15:24:00 \\
5 & 2020-06-09T03:13:00 & 2020-06-09T04:52:00 \\
5 & 2020-06-09T05:45:00 & 2020-06-09T07:36:00 \\
5 & 2020-06-10T05:00:00 & 2020-06-10T23:00:00 \\
5 & 2020-06-11T09:00:00 & 2020-06-12T01:00:00 \\
5 & 2020-06-13T09:00:00 & 2020-06-13T12:00:00 \\
5 & 2020-06-13T16:00:00 & 2020-06-14T01:00:00 \\
\hline
6 & 2020-09-21T04:15:01 & 2020-09-21T07:06:00 \\
6 & 2020-09-21T12:41:00 & 2020-09-21T15:33:00 \\
6 & 2020-09-22T08:13:00 & 2020-09-22T11:54:00 \\
6 & 2020-09-24T21:00:00 & 2020-09-25T08:00:00 \\
6 & 2020-09-25T19:33:00 & 2020-09-25T21:13:00 \\
6 & 2020-09-26T04:26:00 & 2020-09-26T05:09:00 \\
6 & 2020-09-26T06:11:00 & 2020-09-26T08:15:00 \\
6 & 2020-09-26T09:29:00 & 2020-09-26T12:30:00 \\
6 & 2020-09-27T03:06:00 & 2020-09-27T05:15:00 \\
6 & 2020-09-27T10:03:00 & 2020-09-27T11:04:00 \\
6 & 2020-09-27T18:46:00 & 2020-09-27T21:46:00 \\
6 & 2020-09-29T02:04:00 & 2020-09-29T03:49:00 \\
6 & 2020-09-29T08:00:00 & 2020-09-29T10:00:00 \\
6 & 2020-09-29T15:00:00 & 2020-09-29T18:00:00 \\
6 & 2020-09-29T22:00:00 & 2020-09-30T07:00:00 \\
6 & 2020-10-01T00:00:00 & 2020-10-01T06:00:00 \\
6 & 2020-10-01T20:00:00 & 2020-10-02T01:00:00 \\
6 & 2020-10-02T15:00:00 & 2020-10-02T20:00:00 \\
6 & 2020-10-03T06:39:00 & 2020-10-03T09:47:00 \\
\hline
7 & 2021-01-11T09:04:00 & 2021-01-11T10:29:00 \\
7 & 2021-01-11T13:32:00 & 2021-01-11T14:31:00 \\
7 & 2021-01-13T00:00:00 & 2021-01-13T01:00:00 \\
7 & 2021-01-13T08:00:00 & 2021-01-13T11:00:00 \\
7 & 2021-01-14T03:00:00 & 2021-01-14T08:00:00 \\
7 & 2021-01-15T00:00:00 & 2021-01-15T03:00:00 \\
7 & 2021-01-15T18:22:00 & 2021-01-15T22:22:00 \\
7 & 2021-01-16T00:12:00 & 2021-01-16T01:44:00 \\
7 & 2021-01-16T02:09:00 & 2021-01-16T04:15:00 \\
7 & 2021-01-16T05:09:00 & 2021-01-16T07:14:00 \\
7 & 2021-01-16T09:44:00 & 2021-01-16T11:08:00 \\
7 & 2021-01-16T23:02:00 & 2021-01-17T00:16:00 \\
7 & 2021-01-17T09:00:00 & 2021-01-17T13:00:00 \\
7 & 2021-01-17T16:03:00 & 2021-01-17T17:25:00 \\
7 & 2021-01-18T08:47:00 & 2021-01-18T10:08:00 \\
7 & 2021-01-19T01:23:00 & 2021-01-19T02:59:00 \\
7 & 2021-01-19T03:45:00 & 2021-01-19T05:11:00 \\
7 & 2021-01-19T12:39:00 & 2021-01-19T13:24:00 \\
7 & 2021-01-19T23:39:00 & 2021-01-20T07:20:00 \\
7 & 2021-01-21T13:00:00 & 2021-01-21T18:00:00 \\
\hline
8 & 2021-04-24T05:00:00 & 2021-04-24T14:16:00 \\
8 & 2021-04-26T04:24:00 & 2021-04-26T06:31:00 \\
8 & 2021-04-26T11:00:00 & 2021-04-26T12:48:00 \\
8 & 2021-04-26T13:22:00 & 2021-04-26T15:11:00 \\
8 & 2021-04-26T19:28:00 & 2021-04-26T21:11:00 \\
8 & 2021-04-27T15:00:00 & 2021-04-27T17:00:00 \\
8 & 2021-04-28T04:23:00 & 2021-04-28T05:43:00 \\
8 & 2021-04-28T10:00:00 & 2021-04-28T12:00:00 \\
8 & 2021-04-28T13:05:00 & 2021-04-28T16:45:00 \\
8 & 2021-04-28T19:00:00 & 2021-04-28T21:00:00 \\
8 & 2021-04-29T02:01:00 & 2021-04-29T03:32:00 \\
8 & 2021-04-29T06:30:00 & 2021-04-29T08:09:00 \\
8 & 2021-04-29T20:35:00 & 2021-04-30T05:28:00 \\
8 & 2021-05-01T17:41:00 & 2021-05-01T22:24:00 \\
8 & 2021-05-03T04:19:00 & 2021-05-03T06:42:00 \\
\hline
9 & 2021-08-04T20:00:00 & 2021-08-05T08:00:00 \\
9 & 2021-08-05T13:31:00 & 2021-08-05T16:54:00 \\
9 & 2021-08-06T09:31:00 & 2021-08-06T11:58:00 \\
9 & 2021-08-07T14:16:00 & 2021-08-08T02:49:00 \\
9 & 2021-08-08T10:53:00 & 2021-08-08T11:54:00 \\
9 & 2021-08-08T17:20:00 & 2021-08-08T18:43:00 \\
9 & 2021-08-08T21:07:00 & 2021-08-08T22:46:00 \\
9 & 2021-08-09T09:16:00 & 2021-08-09T10:01:00 \\
9 & 2021-08-09T16:52:00 & 2021-08-09T19:56:00 \\
9 & 2021-08-09T21:03:00 & 2021-08-10T00:27:00 \\
9 & 2021-08-10T08:42:00 & 2021-08-10T10:36:00 \\
9 & 2021-08-10T11:37:00 & 2021-08-10T12:29:00 \\
9 & 2021-08-10T18:54:00 & 2021-08-11T01:03:00 \\
9 & 2021-08-11T11:00:00 & 2021-08-11T14:00:00 \\
9 & 2021-08-11T18:00:00 & 2021-08-11T19:00:00 \\
9 & 2021-08-11T20:36:00 & 2021-08-11T22:21:00 \\
9 & 2021-08-11T23:59:00 & 2021-08-12T01:19:00 \\
9 & 2021-08-12T05:39:00 & 2021-08-12T06:45:00 \\
9 & 2021-08-12T19:05:00 & 2021-08-12T20:02:00 \\
9 & 2021-08-14T03:00:00 & 2021-08-14T06:00:00 \\
9 & 2021-08-14T22:59:00 & 2021-08-15T00:15:00 \\
9 & 2021-08-15T01:58:00 & 2021-08-15T03:34:00 \\
\hline
\hline
\end{supertabular}

\section{Discarded intervals}
\label{sec: app_C}
We give in the following table the list of intervals that were discarded in our study during the different encounters. They correspond to either heliospheric current sheet (HCS) crossings, heliospheric plasma sheets (HPS) crossings, coronal mass ejections (CME) or flux ropes, and periods of strahl drop out where magnetic field lines are most likely disconnected from the Sun . All of these intervals are identified visually while scanning through the data. Intervals from encounters 1 to 5 were previously identified in \cite{Fargette_2021}.

\bottomcaption{Timetable of discarded intervals manually selected over encounters 1, 2, 4, 5, 6, 7, 8 and 9}
\begin{supertabular}{cccc}
\hline \hline
Enc & Start time & End time & Tag \\
\hline \hline
1 & 2018-10-31T04:00 &  10-31T12:20 & CME\\
1 & 2018-11-11T17:00 &  11-12T12:00 & CME\\
\hline
4 & 2020-01-30T13:15 &  01-30T17:10 & HPS \\
4 & 2020-01-31T19:50 &  02-01T00:05 & HPS\\
4 & 2020-02-01T03:55 &  02-01T04:15 & HCS \\
\hline
5 & 2020-05-31T12:21 &  06-01T03:40 & Flux ropes\\
5 & 2020-06-01T10:00 &  06-01T16:10 & Strahl inversion\\
5 & 2020-06-01T19:35 &  06-01T21:35 & Flux rope\\
5 & 2020-06-02T06:50 &  06-02T09:10 & HPS\\
5 & 2020-06-04T03:25 &  06-04T06:05 & HPS\\
5 & 2020-06-07T11:10 &  06-07T12:40 & HPS\\
5 & 2020-06-07T20:20 & 06-07T21:10 & HPS \\
5 & 2020-06-08T00:40 & 06-08T12:30 & HCS \\
5 & 2020-06-08T15:30 & 06-09T01:40 & HCS \\
5 & 2020-06-12T01:00 & 06-12T08:00 & Flux rope \\
  & & & or CME \\
\hline
6 & 2020-09-20T11:00 & 09-22T08:00 & multiple HCS \\
6 & 2020-09-25T08:40 & 09-25T19:22 & HCS \\
6 & 2020-09-25T08:40 & 09-25T19:22 & HCS \\
6 & 2020-09-30T09:00 & 09-30T18:00 & probable HCS \\
\hline
7 & 2021-01-17T13:00 & 01-17T15:00 & HCS \\
7 & 2021-01-19T13:24 & 01-19T23:50 & HCS \\
7 & 2021-01-20T07:20 & 01-20T14:00 & compressible  \\
& & & structure \\
7 & 2021-01-22T21:00 & 01-24T12:30 & HCS \\
\hline
8 & 2021-04-23T22:20 & 04-24T04:00 & HCS \\
8 & 2021-04-24T15:57 & 04-24T16:18 & HCS \\
8 & 2021-04-29T00:30 & 04-29T02:01 & HCS \\
8 & 2021-04-29T07:40 & 04-29T10:59 & HCS \\
8 & 2021-04-29T13:38 & 04-29T14:00 & HCS \\
\hline
9 & 2021-08-10T00:27 & 08-10T01:54 & HCS \\
9 & 2021-08-10T10:34 & 08-10T12:06 & HCS \\
9 & 2021-08-10T13:50 & 08-10T18:54 & HCS \\
\hline
\hline
\end{supertabular}

\section{Fitting results}
\label{sec: app_B}

\begin{figure*}
    \centering
    \includegraphics[width=.8\textwidth]{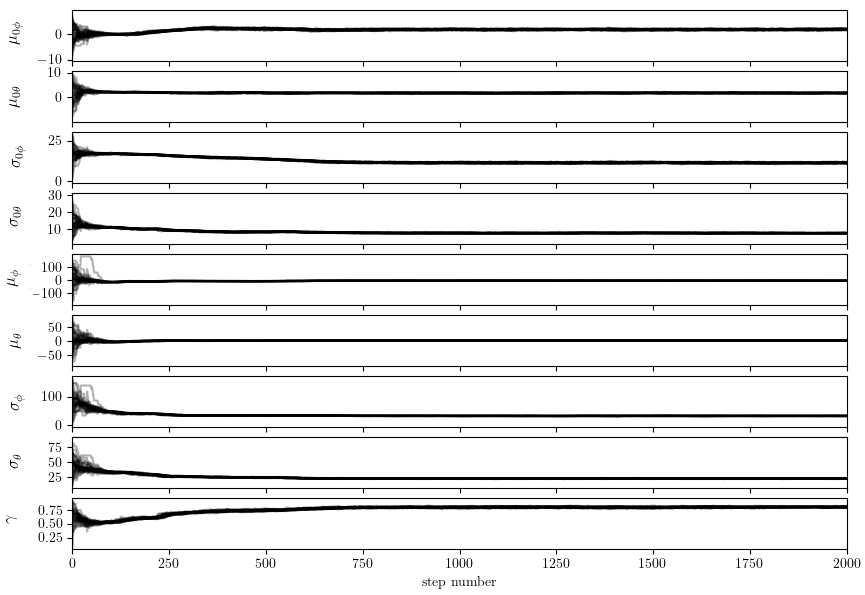}
    \caption{Walker path in the parameter space over 2000 iterations}
    \label{fig: app_fourmis}
\end{figure*}

\begin{figure*}
    \centering
    \includegraphics[width=.8\textwidth]{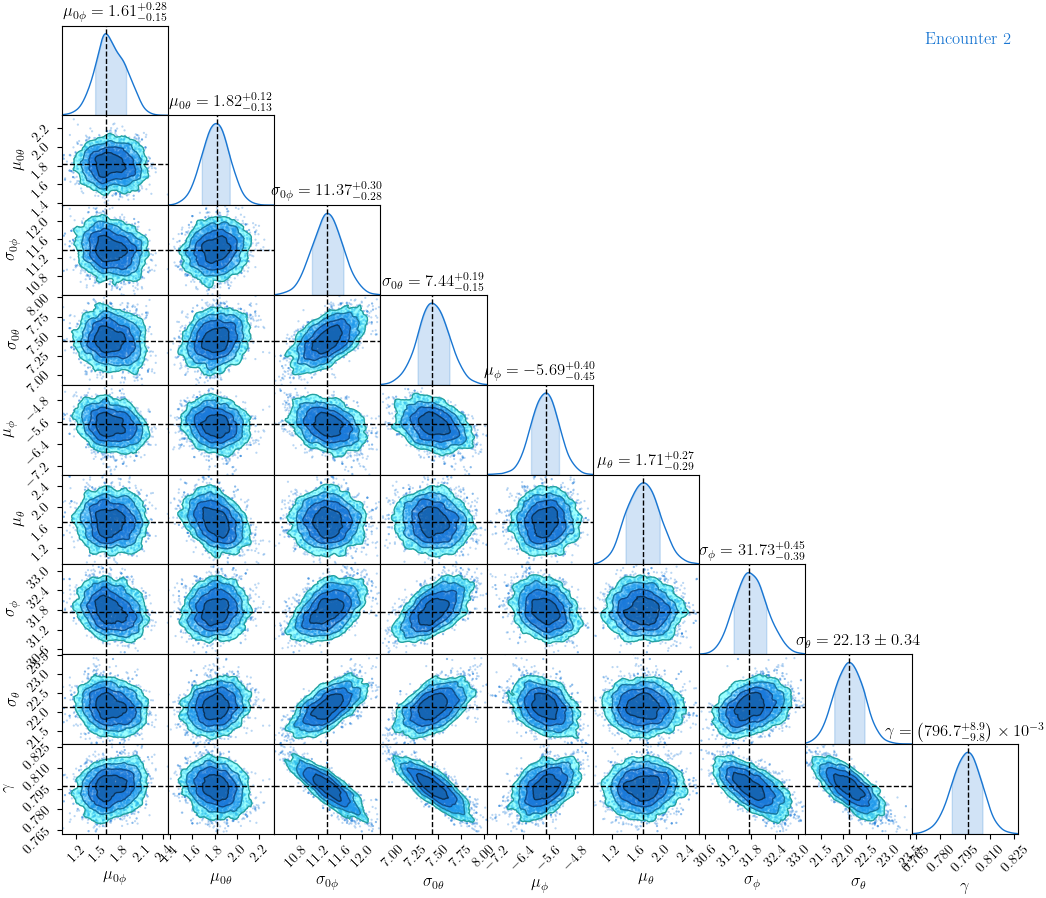}
    \caption{9D probability distribution function of walker positions, discarding the first 1000 iterations.}
    \label{fig: app_mcmc}
\end{figure*}

In section \ref{sec:3_orientation}, we sample the parameter space using the \textit{emcee}\footnote{\href{https://emcee.readthedocs.io/en/stable/}{https://emcee.readthedocs.io/en/stable/}} python library \citep{Foreman-Mackey_2019} which is based on a Monte-Carlo Markov chain algorithm. We use 32 walkers and 2000 iterations, and use the \textit{Chain Consumer} \footnote{\href{https://samreay.github.io/ChainConsumer/chain_api.html}{https://samreay.github.io/ChainConsumer/chain\_api.html}} library to visualise the fitting results.In Figure \ref{fig: app_fourmis} we display the convergence of the fitting algorithm on 2000 iterations. After 1000 steps the results are stable, and so we display the probability distribution function of walker positions in Figure \ref{fig: app_mcmc}, discarding the first 1000 iterations. \footnote{The  associated python code is available here : \href{https://github.com/Nfargette/Fit_double_gaussian/blob/main/Fiting_routine.py}{fit\_double\_gaussian.py}} The fitting results for all encounters are available in table \ref{tab:fit_all_enc}

\begin{table*}
    \centering
    \caption{Most probable (maximum a-posteriori) parameter vectors $\Vec{\mathrm{P}}$, obtained after fitting the double Gaussian model described in the text to the data for all the encounters}
    \label{tab:fit_all_enc}
    \begin{tabular}{c|cccc|cccc|l}
        \hline
		Enc & $\mu_{0\phi}$ & $\mu_{0\theta}$ & $\sigma_{0\phi}$ & $\sigma_{0\theta}$ & $\mu_{\phi}$ & $\mu_{\theta}$ & $\sigma_{\phi}$ & $\sigma_{\theta}$ & $\gamma$ \\ 
		\hline
		
		& & & & & & & & & \\
		
		1 & $0.04^{+0.16}_{-0.20}$ & $1.17^{+0.14}_{-0.16}$ & $12.39^{+0.30}_{-0.38}$ & $10.53\pm 0.24$ & $-5.37^{+0.54}_{-0.38}$ & $-0.06^{+0.28}_{-0.34}$ & $34.77^{+0.66}_{-0.59}$ & $25.01^{+0.41}_{-0.38}$ & $ 0.755^{+0.01}_{-0.002}$  \\
		& & & & & & & & & \\
		2 & $1.61^{+0.28}_{-0.15}$ & $1.82^{+0.12}_{-0.13}$ & $11.37^{+0.30}_{-0.28}$ & $7.44^{+0.19}_{-0.15}$ & $-5.69^{+0.40}_{-0.45}$ & $1.71^{+0.27}_{-0.29}$ & $31.73^{+0.45}_{-0.39}$ & $22.13\pm 0.34$ & $0.797^{+0.009}_{-0.01}$ \\ 
		& & & & & & & & & \\
		
		4 & $0.69^{+0.16}_{-0.14}$ & $1.711^{+0.109}_{-0.099}$ & $14.04^{+0.26}_{-0.17}$ & $10.75^{+0.14}_{-0.15}$ & $-0.89\pm 0.50$ & $0.24^{+0.39}_{-0.44}$ & $34.51^{+0.70}_{-0.69}$ & $24.62^{+0.50}_{-0.46}$ & $0.598^{+0.012}_{-0.017}$ \\ 
		
		& & & & & & & & & \\
		
		5 & $3.85^{+0.18}_{-0.17}$ & $2.60^{+0.11}_{-0.12}$ & $14.23^{+0.20}_{-0.18}$ & $10.19^{+0.16}_{-0.15}$ & $-7.59^{+0.86}_{-0.63}$ & $-1.33^{+0.44}_{-0.63}$ & $33.37^{+0.64}_{-0.66}$ & $26.42^{+0.50}_{-0.45}$ & $0.603\pm 0.012$ \\
		
		& & & & & & & & & \\
		
		6 & $2.11^{+0.15}_{-0.14}$ & $1.642^{+0.096}_{-0.090}$ & $12.46^{+0.22}_{-0.23}$ & $7.99^{+0.15}_{-0.13}$ & $1.79^{+0.37}_{-0.38}$ & $1.46^{+0.25}_{-0.22}$ & $27.80^{+0.47}_{-0.51}$ & $18.50^{+0.41}_{-0.30}$ & $0.660\pm 0.016$ \\ 
		
		& & & & & & & & & \\
		
		7 & $0.51^{+0.14}_{-0.19}$ & $2.778^{+0.098}_{-0.068}$ & $14.22^{+0.16}_{-0.21}$ & $7.56^{+0.14}_{-0.12}$ & $-7.46\pm 0.49$ & $-2.81^{+0.38}_{-0.40}$ & $33.83^{+0.56}_{-0.48}$ & $24.71^{+0.40}_{-0.32}$ & $ 0.688^{+0.0074}_{-0.0094}$ \\ 
		
		& & & & & & & & & \\
		
		8 & $-5.15^{+0.15}_{-0.13}$ & $1.161^{+0.093}_{-0.078}$ & $10.91^{+0.16}_{-0.17}$ & $6.886^{+0.097}_{-0.124}$ & $-11.98^{+0.49}_{-0.56}$ & $-1.76^{+0.30}_{-0.43}$ & $37.35^{+0.57}_{-0.50}$ & $26.71^{+0.46}_{-0.34}$ & $ 0.760^{+0.0055}_{-0.0057}$ \\ 
		
		& & & & & & & & & \\
		
		9 & $-6.57^{+0.16}_{-0.18}$ & $-0.73^{+0.11}_{-0.10}$ & $9.93^{+0.26}_{-0.24}$ & $7.31^{+0.16}_{-0.18}$ & $-9.67^{+0.40}_{-0.45}$ & $-2.75^{+0.32}_{-0.37}$ & $25.29^{+0.58}_{-0.47}$ & $19.66^{+0.41}_{-0.45}$ & $0.707\pm 0.015$ \\ 
		
		& & & & & & & & & \\
		
		\hline
    \end{tabular}
\end{table*}

\end{appendix}
\end{document}